\documentclass[pra,aps,amsmath,amssymb,amsfonts,twocolumn,nofootinbib,floatfix]{revtex4}
\usepackage{}
\usepackage{amssymb}
\usepackage{bm,mathrsfs}
\usepackage{graphicx}
\usepackage{epsfig}
\usepackage{amsmath,bbm}
\usepackage{amsfonts,amssymb}
\usepackage{times}
\usepackage{verbatim}
\usepackage[sort&compress]{natbib}
\usepackage{amsmath}
\usepackage{bm}
\usepackage{float}
\usepackage{textgreek}
\usepackage{textcomp}
\allowdisplaybreaks[4]
\usepackage[colorlinks,breaklinks,linkcolor=blue,anchorcolor=blue,citecolor=blue,urlcolor=magenta]{hyperref}

\usepackage{color}
\definecolor{zzz}{rgb}{0.9,0.0,0.4}

\begin{document}
\title{Quantum Battery Enhancement via Degenerate Optical Parametric Amplifier and Common Reservoirs}
\author{Y. Y. Yang,$^{1}$ H. N. Liu,$^{1}$ Gangcheng Wang,$^{1}$ T. Z. Luan,$^{2}$  and H. Z. Shen$^{1,}$\footnote{\textcolor{zzz}{Corresponding author: shenhz458@nenu.edu.cn }}}
\affiliation{$^1$Center for Quantum Sciences and School of Physics, Northeast Normal University, Changchun 130024, China\\
$^2$College of Physics Science and Technology, Shenyang Normal University, Shenyang 110034, China\\
$^3$Analytical quantum complexity RIKEN Hakubi Research Team,
RIKEN Center for Quantum Computing (RQC), Wako, Saitama 351-0198, Japan\\
$^4$School of Physics and Optoelectronic Engineering, Ludong University, Yantai 264025, China\\
$^5$Quantum Information Research Center and Jiangxi Province Key Laboratory of Applied Optical Technology, Shangrao Normal University, Shangrao 334001, China}

\date{\today}

\begin{abstract}

We investigate a nonreciprocal  quantum battery in which the charger and battery share one or two common  reservoirs, with the charger driven by a degenerate optical parametric amplifier (DOPA). Reservoir-mediated dissipative interactions enable directional energy transfer, while the DOPA provides additional parametric amplification. We show that, in the stable below-threshold regime, the DOPA can enhance the battery energy and average charging power, with the nonlinear gain and pump phase providing effective control of the charging dynamics. Further optimization of the dissipative-coupling asymmetry improves the stored energy. These features persist for both single- and two-common-reservoir configurations, demonstrating a tunable strategy for enhancing open-system quantum-battery charging.

\end{abstract}



\maketitle
\section{Introduction}
Efficient energy transfer via direct interactions between quantum systems is a subject of fundamental and practical interest, with applications ranging from quantum heat engines and thermal logic gates to quantum switches \cite{Wang177208,Joulain200601,Malavazi064146,Campaioli031001,Bai060201,Lobejko260403,Ferraro117702,Andolina047702,Dou023725,Molitor373}. While conventional chemical batteries \cite{Liu028304, Shukla023129, McGilligan044038, GalushkinA1315C7, Zhu19789C98, Shinyama447C52} may intrinsically rely on quantum components, their macroscopic operation remains purely classical. In contrast, quantum mechanical systems capable of storing extractable energy \cite{Strasberg021003, Vinjanampathy545, Alicki180108314, Goold143001, Campisi1653, Gelbwaser329, Giovannetti615, Majer443, Verhagen63, Mari175501, Liao042304, PerDOPAdre134504, Jaynes89, Haroche1083, Schoelkopf664, Chakraborty1999} exhibit genuine quantum effects that can be leveraged to achieve faster and more efficient charging processes than their classical analogs \cite{Horodecki2059, Brouzos062110, Deffner453001, Giovannetti052109, GiovannettiS807}.
The concept of the quantum battery, formally introduced by Alicki and Fannes \cite{Alicki042123}, exemplifies this advantage. Originally proposed as a two-level system for temporarily storing energy transferred from an external field \cite{SenL030402, Pirmoradian043833, Gumberidze19628, Crescente063057, Mohan200614523, Carrega083085, Tabesh052223, Binder075015}, quantum batteries exploit nonclassical phenomena—such as coherence, entanglement, and quantum correlations \cite{Salvia013155,Zhu09401,Shi130602,andolina2025genuine,konar2024quantum,Song020405,Caravelli023095,Arjmandi064106,Cristiano062234,Santos062114,Du013151,Li044118,Rinaldi012205,Ghosh022628,Imai022215,Huang030201,Zhang023187,Song090401,Zhang032211,Arjmandi062609,Dou032212,Song022209,Zheng042442,Khan104318,Kamin052109,Ghosh032207,Seah100601,Zhang140403,Yang012204,Xu012425,Qi032606,Yao044116,Chen052216,Li022217,Andolina205437,Almeida052218,Konar022226,Yang030402,Zakavati054117,Lai023136,Zhao033715}—to potentially surpass classical limits in both charging power and capacity \cite{Campaioli031001}. Following these seminal works \cite{Hovhannisyan240401,Mojaveri064107,Francica062209,Fasihi024117,Wright110302,Zhang042424,Gemme023091,Morrone044073,Yang042205,Shokri064117,Alimuddin022106,Rojo-Francàs032205,Chaki052446,Yao062616,Konar022618,Hu060401,Downing044143,Rodríguez042618,Wang062402,Wang030201,Chen054119,Binninger180202,Wang014121,Arjmandi054115,Liu245418,Tacchino062133,Dong043701,Xu022615,Dias012617,Li054033,Pokhrel130401,Medina220402,Qu180301,Dou115405,Quach024092,Gao043150,Wang042419,Ali052404,Zhang054125,Li052437,Galvão064119,Tirone012204}, a variety of physical substrates have been proposed, including many-body systems \cite{ Rossini115142, TomadinA130, Ito200807089}, spin ensembles \cite{Le022106, Wang205312, Xie034005, Yu054038, Bhattacharjee200807889, Pastur724}, Dicke models \cite{Crescente200909791,Li032409, Zhang181210139}, and the Sachdev-Ye-Kitaev (SYK) model \cite{Ghosh032115,Rosa67, Rossini191207234}. Subsequent research has broadly explored optimal charging protocols \cite{Rodriguez043004,Mazzoncini032218,Crescente033216}, non-Markovian quantum batteries~\cite{Li475614,Xu1092024054132,Zhao1122025024129,Oularabi6792025131003,Khoudiri5652026131178}, topological quantum batteries~\cite{Lu134180401,Zhou042213}, cavity-assisted charging \cite{wang2024cavity,Quach3160}, and open-system dynamics \cite{Kamin022226,Kamin083007,Rodriguez042419,Malavazi16633,Yang109062432,Barra015003,Barra820,Morrone035007}, among other configurations \cite{Glueckstern1541,Carrasco064119,Gyhm140501,Mondal044126,Gemme197,Centrone052213,Dou31503,Yang064069,Abah025201,Song054107,Shaghaghi04LT01,Shaghaghi430,Bakhshinezhad014131,Yang235432,Gemme43,Crescente758,catalano2024frustrating,kurman2026powering,santos2023vacuum,downing2024hyperbolic,liu2024better}. For instance, a notable configuration introduced in Ref.~\cite{Farina035421} utilizes a charger to mediate the coupling between a battery and an external laser field. 

Despite these theoretical advances, maximizing charging efficiency and work extraction \cite{Hovhannisyan033413,Dou43001,Liu18303, Struchtrup250602,  Giorgi035501, GarciaPintos040601, Allahverdyan565, Francica12, Fusco052122, Uzdin124, Monsel130601, Borisenok43,Zhang052106,Julià-Farré023113,Hu042216,Bhattacharjee50} under realistic experimental constraints remains a central challenge.
To address this, reservoir engineering---the controlled design of system--reservoir interactions to steer quantum dynamics \cite{Poyatos4728,Verstraete633,Barreiro486,Murch211,Krauter080503}---offers a robust theoretical framework. Specifically, reservoir-induced nonreciprocity has emerged as an effective mechanism for directional energy transfer \cite{Metelmann133904,Metelmann021025,Toth787,Barzanjeh953,Kerckhoff034002,Chapman041043,Kim063904,Barzanjeh050601}. Recent studies have demonstrated that such nonreciprocal dynamics can significantly enhance the energy transfer efficiency from a charger to a quantum battery \cite{Ahmadi210402,Sun012429,Khan023003,Yang034013}.

In this paper, we investigate the impact of a DOPA on the performance of  quantum batteries coupled to either one or two common reservoirs. The DOPA generates pairs of down-converted photons, thereby introducing single-mode or two-mode squeezing and nonlinear interactions that  enrich the open-system dynamics. By embedding the DOPA within the cavity architecture, we analyze the battery's charging performance as a function of the DOPA gain, phase, and reservoir configuration. We find that the introduction of the DOPA significantly enhances both the stored energy and the charging power in both single- and two-common-reservoir setups.

The paper is organized as follows. In Sec. II, we introduce the physical model and derive the open-system dynamics of the charger–battery system. In Sec. III, we analyze the DOPA-enhanced quantum battery coupled to a single common reservoir, deriving analytical steady-state expressions and identifying optimal dissipation-asymmetry conditions. In Sec. IV, we extend this formalism to the two-common-reservoir scenario to demonstrate the robustness of the DOPA-induced enhancement. Finally, Sec. V summarizes our results and discusses future outlooks, including non-Markovian effects and multi-cell battery networks.
\begin{figure}[h]
\centering{
\includegraphics[width=12.6cm, height=6.5cm, clip]{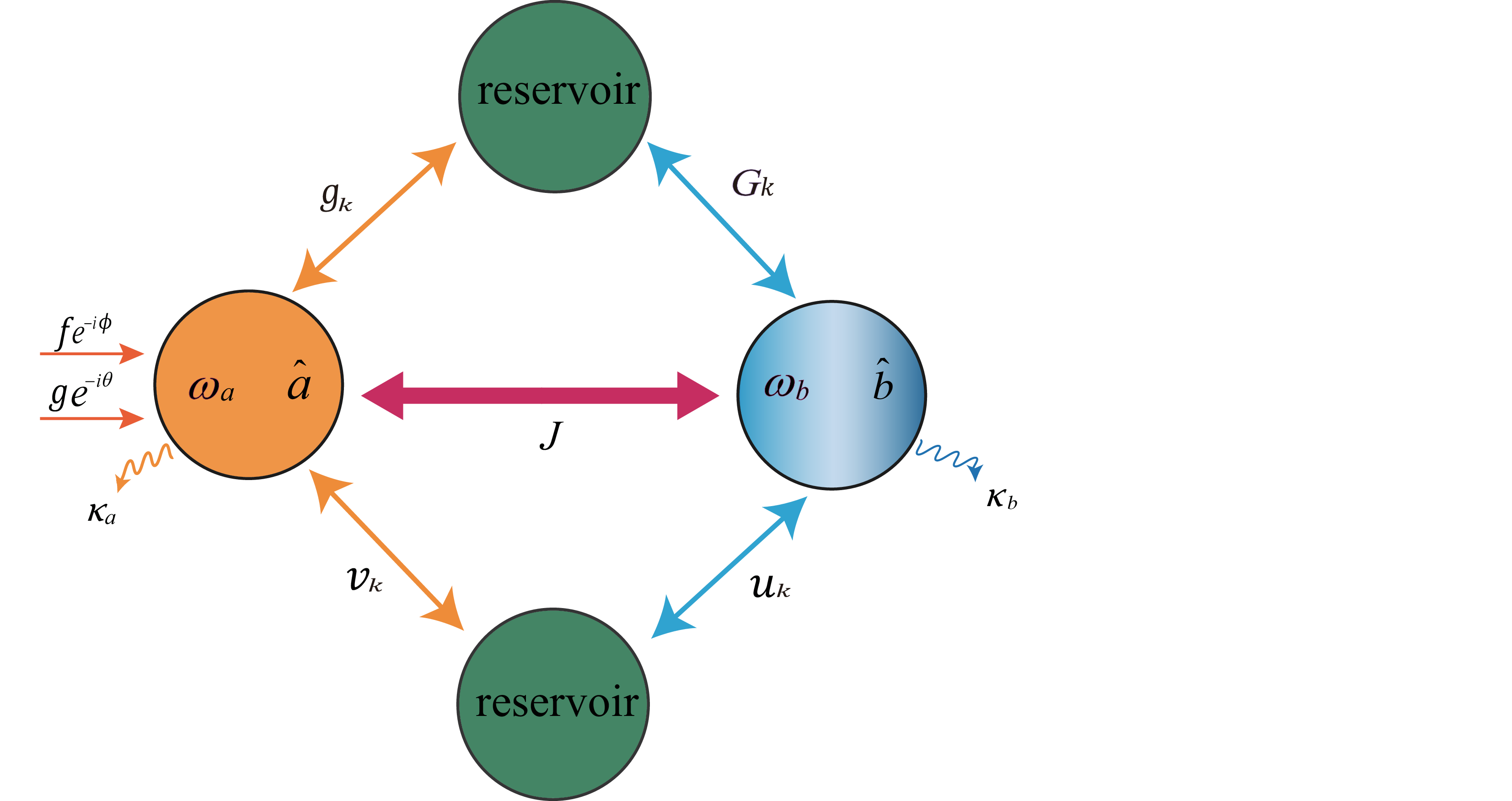}}
\caption{Schematic representation of the enhanced quantum battery system. A quantum charger $a$ coherently interacts with a quantum battery $b$ with a coupling rate $J$. The system is driven by the two-photon pump (strength $g$ and phase $\theta$) and single-photon driving field (strength $f$ and phase $\phi$). The charger and battery are simultaneously coupled to two shared reservoir modes at different rates. The parameters $g_k$ and $v_k$ denote the coupling strengths of the charger to the first and second reservoirs, respectively, while $G_k$ and $u_k$ denote the corresponding coupling strengths of the battery dissirating to reservoirs. In this configuration, energy flows directly from the charger to the battery, while the nonreciprocal condition effectively suppresses energy backflow. Here, $\kappa_a$ and $\kappa_b$ describe the local damping rates of each mode.} \label{model}
\end{figure}

\section{The Model}
The DOPA is a second-order nonlinear optical device that can generate pairs of down-converted photons and exhibit nearly perfect single or dual-mode squeezing \cite{Gerry2005,Li023838,Nation2012,Leghtas347,Clerk1155,Shen023814,Zhou421289,Shen101013826,Shen98023856,Shen109043714,Li260517810,Zhang109033701,Cui5174713,Yang114023703,Zhou113033722}. As illustrated in Fig.~\ref{model}, we consider a quantum battery charging system comprising two coupled harmonic oscillators. The first oscillator serves as the charger with resonance frequency $\omega_a$ and local damping rate $\kappa_a$, while the second oscillator acts as the battery with resonance frequency $\omega_b$ and local damping rate $\kappa_b$. These two modes are coherently coupled with interaction strength $J$. The charger is driven by two mechanisms: a coherent single-photon drive and a DOPA providing parametric amplification. The DOPA introduces a parametric two-photon interaction via spontaneous parametric down-conversion, which enhances the energy charging efficiency. The energy injected into the charger is then transferred to the battery through the charger-battery coupling. The total Hamiltonian governing this system is given by ($\hbar\equiv 1$), $\hat{H}' = \hat{H}_{S} + \hat{H}_{D} + \hat{H}_{E}$, $\hat{H}_S = \omega_a \hat{a}^\dagger \hat{a} + \omega_b \hat{b}^\dagger \hat{b} + (J \hat{a}^\dagger \hat{b} + J^* \hat{b}^\dagger \hat{a}),
\hat{H}_D =F(t)\hat{a}^\dagger + G(t)\hat{a}^{\dagger 2} +F^*(t)\hat{a} + G^*(t)\hat{a}^2,
\hat{H}_E = \sum_k\omega_k \hat{c}_k^\dagger \hat{c}_k + \sum_k\Omega_k \hat{e}_k^\dagger \hat{e}_k
+ i\sum_k (g_k \hat{a} \hat{c}_k^\dagger  - g_k^*\hat{a}^\dagger \hat{c}_k)+ i\sum_k (G_k \hat{b} \hat{c}_k^\dagger  - G_k^*\hat{b}^\dagger \hat{c}_k)
+ i\sum_k (v_k \hat{a} \hat{e}_k^\dagger  - v_k^*\hat{a}^\dagger \hat{e}_k)+ i\sum_k (u_k \hat{b} \hat{e}_k^\dagger  - u_k^*\hat{b}^\dagger \hat{e}_k),$
where $\hat{a}$ and $\hat{b}$ are the annihilation operators of the 
charger and the battery, respectively. 
$\hat{c}_k$ and $\hat{e}_k$ denote the bosonic annihilation operators 
of the $k$th mode of the two reservoirs with eigenfrequencies 
$\omega_k$ and $\Omega_k$, respectively. 
The parameters $g_k$ and $v_k$ represent the coupling rates of 
the charger to the first and second reservoirs, respectively, 
while $G_k$ and $u_k$ denote the corresponding coupling rates 
of the battery to the two reservoirs. 
The single-photon driving field is 
$F(t) = F e^{-i\omega_l t}$ at frequency 
$\omega_l$ with $F = fe^{-i\phi}$, while the two-photon 
pumping field is 
$G(t) = G e^{-i\omega_p t}$ at the pump 
frequency $\omega_p$ with $G = ge^{-i\theta}$. 
Here, $g$ is the strength of the DOPA, 
proportional to the pump field amplitude, and $\theta$ is 
the phase of the DOPA pump field~\cite{Shahidani88053813}. 
We assume that the DOPA is driven by a pump field at 
frequency $\omega_p$~\cite{Liu033822}, such that the signal 
and idler modes are degenerate at frequency 
$\omega_p/2$~\cite{Adiyatullin81329,Nation2012,
Leghtas347,Clerk1155}. To achieve nonreciprocal transmission, we additionally introduce a dissipative interaction between the charger and battery modes. This interaction arises when the two modes are collectively coupled to two common reservoirs, each characterized by a coupling rate. By adiabatically eliminating the reservoir degrees of freedom, an effective dissipative coupling between the charger and battery is established. In a rotating frame defined by $\hat{U}(t) = \exp[-i\omega_l t(\hat{a}^\dagger\hat{a} + \hat{b}^\dagger\hat{b} + \sum_k \hat{c}_k^\dagger \hat{c}_k + \sum_k \hat{e}_k^\dagger \hat{e}_k)]$ with $\omega_p = 2\omega_l$, the total Hamiltonian is transformed to
\begin{equation}
\begin{aligned}
\hat{H} =& \Delta_a \hat{a}^\dagger \hat{a} + \Delta_b \hat{b}^\dagger \hat{b} + J \hat{a}^\dagger \hat{b} + J^* \hat{b}^\dagger \hat{a}\\
& + F\hat{a}^\dagger+F^*\hat{a} + G\hat{a}^{\dagger 2}  + G^*\hat{a}^2 \\
& + \sum\limits_k\alpha_k \hat{c}_k^\dagger \hat{c}_k + \sum\limits_k\beta_k \hat{e}_k^\dagger \hat{e}_k\\
&+ i\sum\limits_k (g_k \hat{a} \hat{c}_k^\dagger  - g_k^*\hat{a}^\dagger \hat{c}_k)+ i\sum\limits_k (G_k \hat{b} \hat{c}_k^\dagger  - G_k^*\hat{b}^\dagger \hat{c}_k)\\
&+ i\sum\limits_k (v_k \hat{a} \hat{e}_k^\dagger  - v_k^*\hat{a}^\dagger \hat{e}_k)+ i\sum\limits_k (u_k \hat{b} \hat{e}_k^\dagger  - u_k^*\hat{b}^\dagger \hat{e}_k),\label{Heff}
\end{aligned}
\end{equation}
where $\Delta = \omega - \omega_l$, $\alpha_k = \omega_k - \omega_l$, and $\beta_k = \Omega_k - \omega_l$ denote the detunings of the system (assuming $\omega_a = \omega_b \equiv \omega$), the $k$th mode of reservoir $c$ (frequency $\omega_k$), and the $k$th mode of reservoir $e$ (frequency $\Omega_k$) from the driving field, respectively. 
The time evolution of the system and reservoir annihilation operators $\hat{a}(t) = \hat{U}^{\dagger}(t)\hat{a}(0)\hat{U}(t)$, $\hat{b}(t) = \hat{U}^{\dagger}(t)\hat{b}(0)\hat{U}(t)$, $\hat{c}_k(t) = \hat{U}^{\dagger}(t)\hat{c}_k(0)\hat{U}(t)$, and $\hat{e}_k(t) = \hat{U}^{\dagger}(t)\hat{e}_k(0)\hat{U}(t)$ with $\hat{U}(t) = e^{-i\hat{H}t}$ are governed by the Heisenberg equation with
\begin{align}
\frac{d\hat{a}}{dt} =&  - (i\Delta+\frac{\kappa_a}{2})\hat{a}-iJ\hat{b}-iF-2iG\hat{a}^\dagger\nonumber\\
&-\sum\limits_k {g_k^*\hat{c}_k}-\sum\limits_k {v_k^*\hat{e}_k},\label{Heq1} \\
\frac{d\hat{b}}{dt} =&  - (i\Delta+\frac{\kappa_b}{2})\hat{b}-iJ^*\hat{a}-\sum\limits_k {G_k^*\hat{c}_k}-\sum\limits_k {u_k^*\hat{e}_k},\label{Heq2}\\
\frac{d\hat{c}_k}{dt} =&  - i\alpha_k\hat{c}_k+ {G_k\hat{b}}+ {g_k\hat{a}},\label{Heq3}\\
\frac{d\hat{e}_k}{dt} =&  - i\beta_k\hat{e}_k+ {u_k\hat{b}}+ {v_k\hat{a}},\label{Heq4}
\end{align}
where the local damping rate ${\kappa _a}/{2}$ and ${\kappa _b}/{2}$ are phenomenologically added in the above equations. Through a simple calculation by solving Eqs.~(\ref{Heq3}) and (\ref{Heq4}), we obtain the formal solution of the reservoir operators $\hat{c}_k = \hat{c}_k(0)e^{-i\alpha_k t} + g_k \int_0^t d\tau\,\hat{a}(\tau)e^{-i\alpha_k(t-\tau)} + G_k \int_0^t d\tau\,\hat{b}(\tau)e^{-i\alpha_k(t-\tau)}$, $\hat{e}_k = \hat{e}_k(0)e^{-i\beta_k t} + v_k \int_0^t d\tau\,\hat{a}(\tau)e^{-i\beta_k(t-\tau)} + u_k \int_0^t d\tau\,\hat{b}(\tau)e^{-i\beta_k(t-\tau)}$,  where the first terms on the right-hand sides of the equations for $\hat{c}_k$ and $\hat{e}_k$ represent the freely evolving components of the reservoir fields. The second terms account for the backaction of the charger on the reservoirs. The third terms account for the backaction of the battery. Substituting the above $\hat{c}_k$ and $\hat{e}_k$  into Eqs.~\eqref{Heq1} and \eqref{Heq2}, we can obtain the integrodifferential equations

\allowdisplaybreaks[4]
\begin{align}
\frac{d\hat{a}}{dt} ={}&  - (i\Delta+\frac{\kappa_a}{2})\hat{a}-iJ\hat{b}-iF-2iG\hat{a}^\dagger-\hat{K}_1(t)-\hat{K}_3(t)\nonumber\\
&-\int_0^t {d\tau }  \hat{a}(\tau )f_1(t-\tau)-\int_0^t {d\tau }  \hat{a}(\tau )f_3(t-\tau) \nonumber\\
&-\int_0^t {d\tau }  \hat{b}(\tau )f_5(t-\tau)-\int_0^t {d\tau }  \hat{b}(\tau )f_7(t-\tau),\nonumber\\
\frac{d\hat{b}}{dt} ={}&  - (i\Delta+\frac{\kappa_b}{2})\hat{b}-iJ^*\hat{a}-\hat{K}_2(t)-\hat{K}_4(t)\nonumber\\
&-\int_0^t {d\tau }  \hat{a}(\tau )f_6(t-\tau)-\int_0^t {d\tau }  \hat{a}(\tau )f_8(t-\tau) \nonumber\\
&-\int_0^t {d\tau }  \hat{b}(\tau )f_2(t-\tau)-\int_0^t {d\tau }  \hat{b}(\tau )f_4(t-\tau).\label{wentaiji}
\end{align}
\allowdisplaybreaks[0]where the externally driven reservoirs operator $\hat{K}_1(t)$=$\sum\nolimits_k {g_k^*\hat{c}_k(0){e^{ - i{\alpha _k}t}}}$, $\hat{K}_2(t)$=$\sum\nolimits_k {G_k^*\hat{c}_k(0){e^{ - i{\alpha _k}t}}}$, $\hat{K}_3(t)$=$\sum\nolimits_k {v_k^*\hat{e}_k(0){e^{ - i{\beta _k}t}}}$, $\hat{K}_4(t)$=$\sum\nolimits_k {u_k^*\hat{e}_k(0){e^{ - i{\beta _k}t}}}$. We have made the replacements ${g_k}\to g(\omega' )$, ${G_k}\to G(\omega' )$, ${v_k}\to v(\omega' )$, ${u_k}\to u(\omega' )$.
The correlation functions are given by $f_j(t) = \int {J_j(\omega' ){e^{ - i(\omega' - \omega_l) t}}d\omega' }$, where $j$=1-8. We show that nonreciprocity is achieved by balancing the dissipative and coherent
coupling rates. In Eq.~\eqref{wentaiji}, by imposing the condition
$v_{k}^{*}u_{k} = -iJ/\pi - g_{k}^{*}G_{k}$,
the mode $\hat{a}$ becomes completely decoupled from $\hat{b}$,
establishing a unidirectional energy flow from the charger to the battery. $J_1(\omega' ) = \sum\nolimits_k {|{g_k}{|^2}} \delta (\omega'  - {\omega _k})$,
 $J_2(\omega' ) = \sum\nolimits_k {|{G_k}{|^2}} \delta (\omega'  - {\omega _k})$, 
 $J_3(\omega' ) = \sum\nolimits_k {|{v_k}{|^2}} \delta (\omega'  - {\Omega _k})$, 
$J_4(\omega' ) = \sum\nolimits_k {|{u_k}{|^2}} \delta (\omega'  - {\Omega _k})$, 
 $J_5(\omega' ) = \sum\nolimits_k {{g^*_k}{G_k}} \delta (\omega'  - {\omega _k})$, 
$J_6(\omega' ) = \sum\nolimits_k {{G^*_k}{g_k}} \delta (\omega'  - {\omega _k})$, 
$J_7(\omega' ) = \sum\nolimits_k {{v^*_k}{u_k}} \delta (\omega'  - {\Omega _k})$, and $J_8(\omega' ) = \sum\nolimits_k {{u^*_k}{v_k}} \delta (\omega'  - {\Omega _k})$ denote the spectral densities of the reservoirs.
\section{DOPA Effects in a Single Common Reservoir Configuration}
\subsection{Differential equation of motion and analytical solution of steady-state energy}
\begin{figure}[t]
\centerline{
\includegraphics[width=0.46\textwidth]{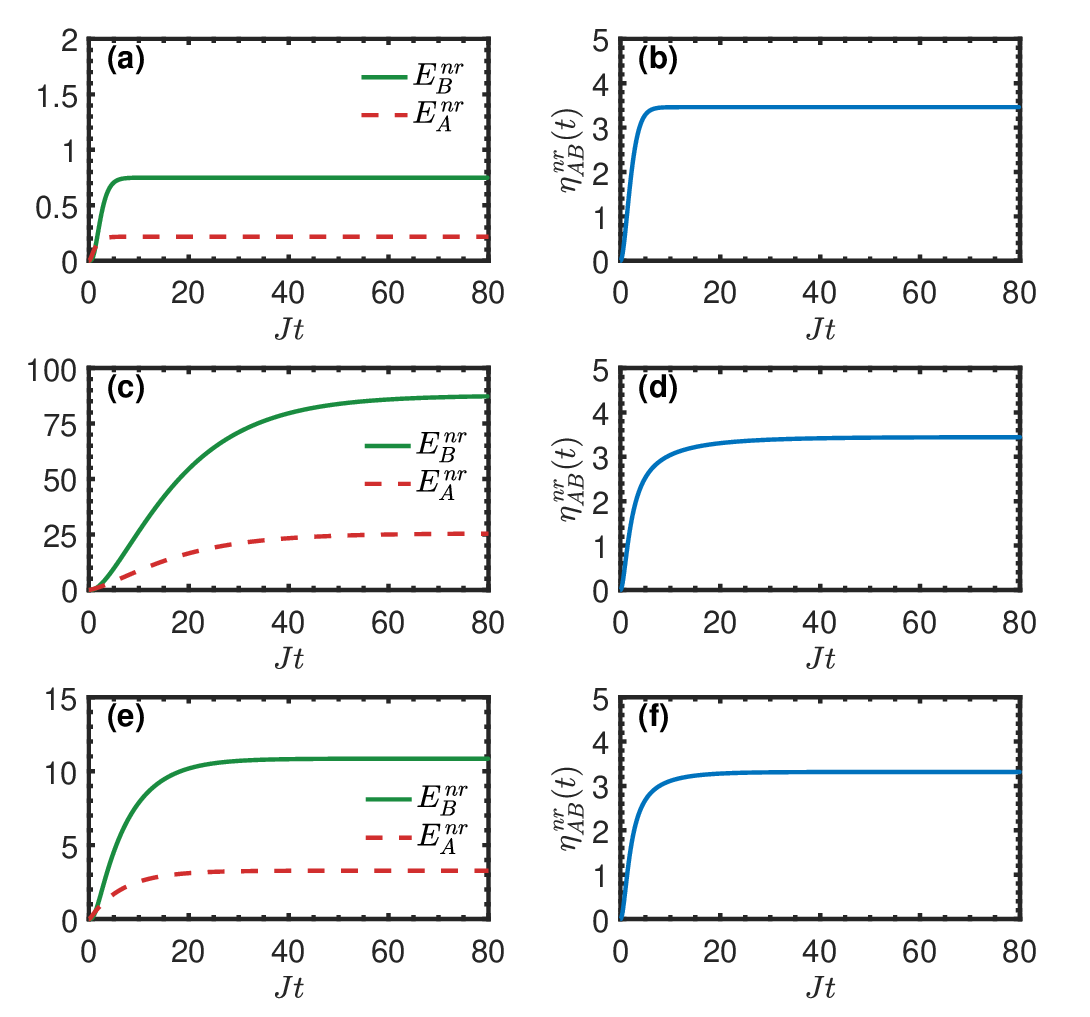}}
\caption{Dynamics of the charger energy $E_A^{\text{nr}}$, the 
    battery energy $E_B^{\text{nr}}$, and their ratio 
    $\eta_{AB}^{\text{nr}}(t) = E_B^{\text{nr}}/E_A^{\text{nr}}$ 
    as functions of the scaled time $Jt$ in the nonreciprocal regime. 
    The DOPA nonlinear gain $g$ and pump-field phase $\theta$ are set 
    to: (a) and (b) $g = 0$, $\theta = 0$; (c) and (d) 
    $g = 0.1\omega$, $\theta = 0$; (e) and (f) $g = 0.1\omega$, 
    $\theta = 3\pi/2$. 
    The other parameters are ${\kappa_a} ={\kappa_b} = 0.03{\omega}$, ${\Gamma _1} ={\Gamma _2}=0.4{\omega}$, $f = 0.01\omega$, $\phi=0$, and $ {|J|} =0.2 \omega$. \label{energy_a_b.eps}}
\end{figure}

\begin{figure}[t]
\centerline{
\includegraphics[width=0.46\textwidth]{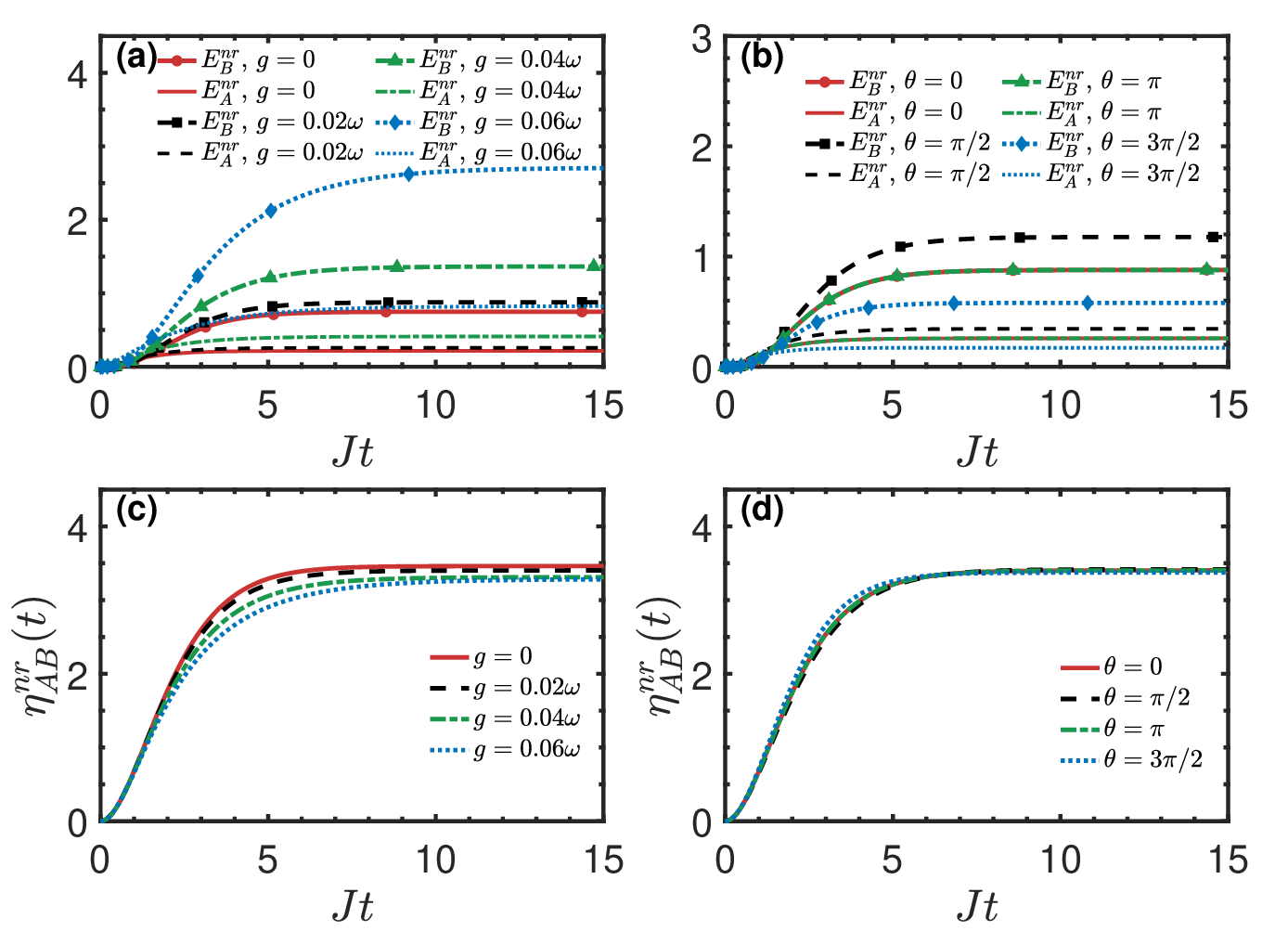}}
\caption{Dynamics of the charger energy $E_A^{\text{nr}}$, the 
    battery energy $E_B^{\text{nr}}$, and their ratio 
    $\eta_{AB}^{\text{nr}}(t) = E_B^{\text{nr}}/E_A^{\text{nr}}$ 
    as functions of the scaled time $Jt$ in the nonreciprocal regime. 
    (a) and (c) show results for different values of the DOPA nonlinear 
    gain $g$ at a fixed pump-field phase $\theta = 0$. 
    (b) and (d) show results for different values of the pump-field 
    phase $\theta$ at a fixed gain $g = 0.02\omega$.
The other parameters are the same as those in Fig.~\ref{energy_a_b.eps}.} \label{energy_a_b_theta_g.eps}
\end{figure}

\begin{figure}[h]
\centering
\includegraphics[width=0.4\textwidth]{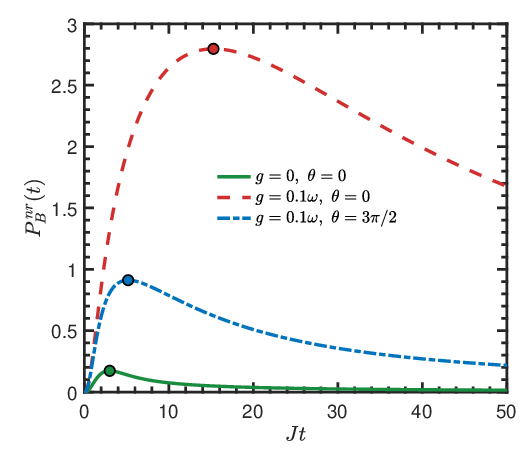}
\caption{Average charging power $P_B^{\text{nr}}$ as a function of the scaled time $Jt$ in the nonreciprocal case. The other parameters are the same as in Fig.~\ref{energy_a_b.eps}.}
\label{thesis3/Tex/p_nr.png}
\end{figure}

We first consider the case where the charger and the battery are coupled to a single common reservoir. Under the Markov approximation, the spectral response functions are taken as $g(\omega') = \sqrt{{\Gamma_1}/{2\pi}},
G(\omega') = \sqrt{{\Gamma_2}/{2\pi}},v(\omega') = 0,u(\omega') =0$.
 The spectral densities of the reservoirs are 
$J_1(\omega') = {\Gamma_1}/{2\pi},
J_2(\omega') = {\Gamma_2}/{2\pi},
J_3(\omega') = J_4(\omega') = J_7(\omega') = J_8(\omega') = 0,J_5(\omega')= J_6(\omega') = {\sqrt{\Gamma_1 \Gamma_2}}/{2\pi}$. The dynamics of the system is governed by the following coupled differential equations for the first- and second-order moments
\begin{small}
\begin{align}
\frac{d\hat{a}}{dt} ={}& -(i\Delta + \frac{\Lambda_1}{2})\hat{a}
    - iF - 2iG\hat{a}^\dagger
    - (iJ + \frac{\sqrt{\Gamma_1\Gamma_2}}{2})\hat{b}\nonumber\\
    &-\hat{K}_1(t),\nonumber\\
\frac{d\hat{b}}{dt} ={}& -(i\Delta + \frac{\Lambda_2}{2})\hat{b}
    - (iJ^* + \frac{\sqrt{\Gamma_1\Gamma_2}}{2})\hat{a}-\hat{K}_2(t),\nonumber\\
\frac{d\hat{a}^\dagger\hat{a}}{dt} ={}& -\Lambda_1\hat{a}^\dagger\hat{a}
    - iF\hat{a}^\dagger + iF^*\hat{a}
    - 2iG\hat{a}^{\dagger 2} + 2iG^*\hat{a}^2\nonumber\\
    &- (iJ + \frac{\sqrt{\Gamma_1\Gamma_2}}{2})\hat{a}^\dagger\hat{b}
    + (iJ^* - \frac{\sqrt{\Gamma_1\Gamma_2}}{2})\hat{b}^\dagger\hat{a}\nonumber\\
    &-\hat{a}^\dagger\hat{K}_1(t)-\hat{K}^\dagger_1(t)\hat{a},\nonumber\\
\frac{d\hat{b}^\dagger\hat{b}}{dt} ={}& -\Lambda_2\hat{b}^\dagger\hat{b}-\hat{b}^\dagger\hat{K}_2(t)-\hat{K}^\dagger_2(t)\hat{b}\nonumber\\
    &+ (iJ - \frac{\sqrt{\Gamma_1\Gamma_2}}{2})\hat{a}^\dagger\hat{b}
    - (iJ^* + \frac{\sqrt{\Gamma_1\Gamma_2}}{2})\hat{b}^\dagger\hat{a},\nonumber\\
\frac{d\hat{a}^\dagger\hat{b}}{dt} ={}& -\frac{\Lambda_1+\Lambda_2}{2}\,\hat{a}^\dagger\hat{b}
    - (iJ^* + \frac{\sqrt{\Gamma_1\Gamma_2}}{2})\hat{a}^\dagger\hat{a}\nonumber\\
    &+ (iJ^* - \frac{\sqrt{\Gamma_1\Gamma_2}}{2})\hat{b}^\dagger\hat{b}
    + iF^*\hat{b} + 2iG^*\hat{a}\hat{b}\nonumber\\
    &-\hat{a}^\dagger\hat{K}_2(t)-\hat{K}^\dagger_1(t)\hat{b},\nonumber\\
\frac{d\hat{a}^2}{dt} ={}& -(2i\Delta + \Lambda_1)\hat{a}^2
    - (2iJ + \sqrt{\Gamma_1\Gamma_2})\hat{a}\hat{b}
    - 2iF\hat{a}\nonumber\\
    &- 4iG\hat{a}^\dagger\hat{a}-2iG-2\hat{a}\hat{K}_1(t),\nonumber\\
\frac{d\hat{b}^2}{dt} ={}& -(2i\Delta + \Lambda_2)\hat{b}^2
    - (2iJ^* + \sqrt{\Gamma_1\Gamma_2})\hat{a}\hat{b}\nonumber\\
    &-2\hat{b}\hat{K}_2(t),\nonumber\\
\frac{d\hat{a}\hat{b}}{dt} ={}& -(2i\Delta + \frac{\Lambda_1+\Lambda_2}{2})\hat{a}\hat{b}
    - iF\hat{b} - 2iG\hat{a}^\dagger\hat{b}\nonumber\\
    &- (iJ + \frac{\sqrt{\Gamma_1\Gamma_2}}{2})\hat{b}^2
    - (iJ^* + \frac{\sqrt{\Gamma_1\Gamma_2}}{2})\hat{a}^2\nonumber\\
    &-\hat{a}\hat{K}_2(t)-\hat{K}_1(t)\hat{b},\label{twohaishenbao}
\end{align}
\end{small}where $\Lambda_1=\Gamma_1+\kappa_a$ and $\Lambda_2=\Gamma_2+\kappa_b$ denote the effective dissipation rates of the charger and the battery, respectively, incorporating both the intrinsic dissipation rates $\kappa_a$ ($\kappa_b$) of the corresponding resonators and the additional dissipation rates $\Gamma_1$ ($\Gamma_2$) induced by the common reservoir. When the system satisfies the nonreciprocal condition
$J=i\sqrt{\Gamma_1\Gamma_2}/2$, the term
$\bigl(iJ+\sqrt{\Gamma_1\Gamma_2}/2\bigr)\hat{b}$
in the equation for $d\hat{a}/dt$ in Eq.~\eqref{twohaishenbao} vanishes. Consequently, the backaction of the battery on the charger is completely eliminated, resulting in unidirectional energy transfer from the charger to the battery. Assuming that both the charger and the battery are initially in their vacuum states, we impose the initial conditions$\langle \hat{b}\rangle
= \langle \hat{b}^\dagger\rangle
= \langle \hat{a}\rangle
= \langle \hat{a}^\dagger\rangle
= \langle \hat{b}^\dagger \hat{b}_j\rangle
= 0, \langle \hat{b}\hat{b}_j\rangle= \langle \hat{b}^\dagger\hat{b}_j^\dagger\rangle
= \langle \hat{a}\hat{b}^\dagger\rangle
= \langle \hat{a}^\dagger\hat{b}\rangle
= 0, \langle \hat{a}\hat{b}\rangle
= \langle \hat{a}^\dagger\hat{b}^\dagger\rangle= \langle \hat{a}^\dagger\hat{a}\rangle
= \langle \hat{a}^2\rangle
= \langle \hat{a}^{\dagger 2}\rangle
= 0$. To obtain a closed set of equations of motion for the expectation values of the system operators, we further assume that the system and the environment are initially uncorrelated, such that the total density operator can be factorized as
$\rho_{\rm tot}(0)=\rho_{ab}(0)\otimes\rho_E$.
The initial state of the system is taken as the joint vacuum state of the charger and the battery $\rho_{ab}(0)=|0\rangle_a\langle0|\otimes|0\rangle_b\langle0|$,
while the common reservoir is assumed as in the zero-temperature vacuum state
$\rho_E=|0\rangle_E\langle0|$.
Under these assumptions, the environmental noise operators have vanishing expectation values $\langle\hat{K}_i(t)\rangle=0$ for $i=1,2$. Moreover, since the system and the environment are initially uncorrelated, we have$\langle \hat{O}_S(0)\hat{K}_i(t)\rangle
=
\langle \hat{K}_i^\dagger(t)\hat{O}_S(0)\rangle
=0$, where $\hat{O}_S$ denotes an arbitrary system operator. Therefore, upon taking the expectation values of Eq.~\eqref{twohaishenbao}, all terms containing a single environmental noise operator give no contribution to the averaged system dynamics.  On the other hand, the vacuum-fluctuation contribution induced by the DOPA is already encoded in the  commutation relations. Consequently, the inhomogeneous term $-2iG$ must be retained in the equation of motion for $\langle\hat{a}^2\rangle$. Under the above assumptions of a zero-temperature Markovian vacuum reservoir,  the evolution of the expectation values of the operators over time is
\begin{small}
\begin{align}
\frac{d\langle\hat{a}\rangle}{dt}=&-(i\Delta+\frac{\Lambda_1}{2})\langle\hat{a}\rangle-iF-2iG\langle\hat{a}^\dagger\rangle,\nonumber\\
\frac{d\langle\hat{b}\rangle}{dt}=&-(i\Delta+\frac{\Lambda_2}{2})\langle\hat{b}\rangle-\sqrt{\Gamma_1\Gamma_2}\langle\hat{a}\rangle,\nonumber\\
\frac{d\langle\hat{a}^\dagger\hat{a}\rangle}{dt}=&-\Lambda_1\langle\hat{a}^\dagger\hat{a}\rangle-iF\langle\hat{a}^\dagger\rangle+iF^*\langle\hat{a}\rangle\nonumber\\
&-2iG\langle\hat{a}^{\dagger 2}\rangle+2iG^*\langle\hat{a}^2\rangle,\nonumber\\
\frac{d\langle\hat{b}^\dagger\hat{b}\rangle}{dt}=&-\Lambda_2\langle\hat{b}^\dagger\hat{b}\rangle-\sqrt{\Gamma_1\Gamma_2}\langle\hat{a}^\dagger\hat{b}\rangle-\sqrt{\Gamma_1\Gamma_2}\langle\hat{b}^\dagger\hat{a}\rangle,\nonumber\\
\frac{d\langle\hat{a}^\dagger\hat{b}\rangle}{dt}=&-\frac{(\Lambda_1+\Lambda_2)}{2}\langle\hat{a}^\dagger\hat{b}\rangle-\sqrt{\Gamma_1\Gamma_2}\langle\hat{a}^\dagger\hat{a}\rangle\nonumber\\
&+iF^*\langle\hat{b}\rangle+2iG^*\langle\hat{a}\hat{b}\rangle,\nonumber\\
\frac{d\langle\hat{a}^2\rangle}{dt}=&-(2i\Delta+\Lambda_1)\langle\hat{a}^2\rangle-2iF\langle\hat{a}\rangle-4iG\langle\hat{a}^\dagger\hat{a}\rangle-2iG,\nonumber\\
\frac{d\langle\hat{a}\hat{b}\rangle}{dt}=&-(2i\Delta+\frac{\Lambda_1+\Lambda_2}{2})\langle\hat{a}\hat{b}\rangle-\sqrt{\Gamma_1\Gamma_2}\langle\hat{a}^2\rangle\nonumber\\
&-iF\langle\hat{b}\rangle-2iG\langle\hat{a}^\dagger\hat{b}\rangle.\nonumber\\
\label{second1}
\end{align}
\end{small}To quantitatively characterize the energy storage performance of the system, the energies stored in the charger and the battery during the charging process are defined as~\cite{AlickiL103}, $E_j(t) = \omega\,\mathrm{Tr}\bigl[\rho_j(t)\,\hat{j}^\dagger\hat{j}\bigr]
        = \omega\bigl\langle\hat{j}^\dagger(t)\hat{j}(t)\bigr\rangle, \quad j = a,\, b$. Taking the steady-state limit ($t \to \infty$) of the equations of motion under the resonance condition $\Delta_{a} = \Delta_{b} = 0$, we obtain analytical expressions for the steady-state energies of the charger and battery in the nonreciprocal regime
\begin{small}
\begin{align}
{E_B^{nr}(\infty)}=&\frac{128 \omega |J|^2|G|^2(2\Lambda_1+\Lambda_2)}
{\Lambda_2(16|G|^2-\Lambda_1^2)
[16|G|^2-(\Lambda_1+\Lambda_2)^2]}\nonumber\\
&+\frac{64 \omega |J|^2 |F|^2(16|G|^2+\Lambda_1^2)}{(16|G|^2-\Lambda_1^2)^2\Lambda_2^2}\nonumber\\
&+
\frac{256 i \omega |J|^2 \Lambda_1
[(F^*)^2G-F^2G^*]}
{(16|G|^2-\Lambda_1^2)^2\Lambda_2^2}
,\label{wenenergyb}\\
E_A^{nr}(\infty)
={}&
\frac{
64\omega |G|^2(|F|^2-2|G|^2)
}{
(-16|G|^2+\Lambda_1^2)^2
}
\nonumber\\
&+
\frac{
16i\omega[(F^*)^2G-F^2G^*]\Lambda_1
}{
(-16|G|^2+\Lambda_1^2)^2
}
\nonumber\\
&+
\frac{
4\omega(|F|^2+2|G|^2)\Lambda_1^2
}{
(-16|G|^2+\Lambda_1^2)^2
},\label{wenenergya}
\end{align}
\end{small}where the superscript $nr$ labels the nonreciprocal regime. As shown by Eq.~\eqref{wenenergya}, the steady-state energy
$E_{A}^{{nr}}(\infty)$ of the charger is determined solely by
the driving parameters $F$, $G$, and the effective dissipation rate
$\Lambda_{1}$ of the charger itself, and is completely independent of
the effective dissipation rate $\Lambda_{2}$ of the battery.
This is consistent with the nature of nonreciprocal dynamics, in which
the charger is unaffected by any backaction from the battery.
Eq.~\eqref{wenenergyb} further shows that the steady-state energy
$E_{B}^{{nr}}(\infty)$ of the battery depends on both
$\Lambda_{1}$ and $\Lambda_{2}$, consistent with the unidirectional
energy transfer from the charger to the battery.
We emphasize that the steady-state solutions in Eqs.~\eqref{wenenergya} and ~\eqref{wenenergyb} exist only in the stable below-threshold regime of the DOPA. In the nonreciprocal limit, the charger dynamics is decoupled from the battery backaction, and the relevant drift matrix for $(\hat a,\hat a^\dagger)^T$ is
\begin{equation}
\mathcal{M}_a=
\begin{pmatrix}
-\Lambda_1/2-i\Delta & -2iG \\
2iG^* & -\Lambda_1/2+i\Delta
\end{pmatrix},
\end{equation}
whose eigenvalues are $\lambda_{\pm}
=-{\Lambda_1}/{2}
\pm
\sqrt{4|G|^2-\Delta^2}$. The stability condition ${\rm Re}\lambda_{\pm}<0$ therefore gives $16|G|^2<\Lambda_1^2+4\Delta^2 $.
For the resonant case $\Delta=0$, this reduces to the standard below-threshold condition $4|G|<\Lambda_1 $.
When this condition is violated, the parametric gain exceeds the effective damping of the charger mode and no finite steady state exists within the present Markovian model.
In the limit of symmetric damping rates, i.e., $\Lambda = \Lambda_1 = \Lambda_2$ (which requires $\Gamma = \Gamma_1 = \Gamma_2$ and $\kappa_a = \kappa_b$), Eqs.~\eqref{wenenergyb} and ~\eqref{wenenergya} can be simplified to
\begin{small}
\begin{align}
{E_B^{nr}(\infty)}=&\frac{96 \omega |J|^2|G|^2 }
{(16|G|^2-\Lambda^2)
[4|G|^2-\Lambda^2]}\nonumber\\
&+\frac{64 \omega |J|^2 |F|^2(16|G|^2+\Lambda^2)}{(16|G|^2-\Lambda^2)^2\Lambda^2}\nonumber\\
&+
\frac{256 i \omega |J|^2 \Lambda
[(F^*)^2G-F^2G^*]}
{(16|G|^2-\Lambda^2)^2\Lambda^2},\label{wenenergybb}\\
E_A^{nr}(\infty)
={}&\frac{64\omega |G|^2(|F|^2-2|G|^2)
}{(-16|G|^2+\Lambda^2)^2}\nonumber\\
&+\frac{
16i\omega[(F^*)^2G-F^2G^*]\Lambda
}{(-16|G|^2+\Lambda^2)^2}\nonumber\\
&+\frac{
4\omega(|F|^2+2|G|^2)\Lambda^2
}{
(-16|G|^2+\Lambda^2)^2
} ,\label{wenenergyaa}
\end{align}
\end{small}which allow us to further elucidate the synergistic enhancement mechanism arising from the interplay between DOPA and nonreciprocity. Substituting $F = f\,e^{-i\phi}$ and $G = g\,e^{-i\theta}$
into Eq.~(\ref{wenenergybb}), we can express the steady-state
energy of the battery as
\begin{small}
\begin{align}
E_B^{nr}(\infty)={}&\frac{32\omega |J|^2(-48g^4\Lambda^2+3g^2\Lambda^4
)}{(-4g^2+\Lambda^2)(-16g^2\Lambda+\Lambda^3)^2
}\notag\\
&+\frac{64\omega f^2|J|^2(-64g^4+12g^2\Lambda^2+\Lambda^4)}{(-4g^2+\Lambda^2)(-16g^2\Lambda+\Lambda^3)^2}\notag\\
&+\frac{512\omega f^2|J|^2g\Lambda\sin\theta'
}{(-16g^2\Lambda+\Lambda^3)^2},\label{eq:EB_phase}
\end{align}
\end{small}where $\theta' = \theta - 2\phi$ denotes the relative phase between the two-photon pump and the coherent drive. From Eq.~(\ref{eq:EB_phase}), the dependence of the battery
energy storage on the DOPA pump-field phase $\theta'$ becomes
evident.
When $\sin\theta' = 1$ (e.g., $\theta = \pi/2$, $\phi = 0$),
the interference term reaches its maximum positive value, while
the parametric amplification interferes constructively with the
driving field, leading to peak battery energy storage.
When $\sin\theta' = 0$ ($\theta' = 0$ or $\pi$), the
interference term vanishes, where the stored energy is determined
solely by the phase-independent terms.
When $\sin\theta' = -1$ ($\theta' = 3\pi/2$), the interference
term reaches its maximum negative value, producing a destructive
effect that minimizes the stored energy.
We emphasize that Eqs.~\eqref{wenenergybb} and \eqref{wenenergyaa} are valid only in the stable below-threshold regime. We next analyze the energy distribution between the charger and
the battery.
The energy ratio is defined as
$\eta_{AB}^{\text{nr}}(t) = E_B^{\text{nr}}(t)/E_A^{\text{nr}}(t)$.
Combining Eqs.~(\ref{wenenergybb}) and (\ref{wenenergyaa})
and eliminating the driving parameters, we obtain the
closed-form expression for the steady-state energy ratio
$\eta_{AB}^{\text{nr}}(\infty) = S_d$.
This result shows that the nonreciprocal mechanism establishes a
unidirectional dissipative channel from the charger to the
battery via asymmetric coupling to the common reservoir.
The transmission efficiency is characterized by the
proportionality factor
\begin{equation}\label{eq:Sd_phase}
\begin{aligned}
S_d={}&\frac{16|J|^2}{\Lambda^2}+\frac{8|J|^2(256g^6-32g^4\Lambda^2)}{\mathcal{D}}+\frac{8|J|^2g^2\Lambda^4}{\mathcal{D}},
\end{aligned}
\end{equation}
where $\mathcal{D} = \Lambda^2(4g^2-\Lambda^2)[-32g^4+2g^2\Lambda^2+f^2(16g^2+\Lambda^2)+8f^2g\Lambda\sin\theta']$.
It can be seen that the nonreciprocal mechanism establishes a unidirectional energy-transfer channel from the charger to the battery through the common reservoir, with the transfer efficiency characterized by the proportionality factor $S_d$. Efficient energy storage in the battery requires $S_d>1$.
Figure~\ref{energy_a_b.eps} shows the time evolution of the energy stored in the charger $E_A^{\text{nr}}$, the energy stored in the battery $E_B^{\text{nr}}$, and their ratio $\eta_{AB}^{\text{nr}}(t)$ for the single-common-reservoir-coupled quantum battery under the nonreciprocal condition. In the absence of the DOPA ($g=0$, $\theta=0$), the energy initially stored in the battery is lower than that in the charger. As the unidirectional energy transfer mediated by the nonreciprocal dissipative channel proceeds, the battery energy gradually increases, exceeds the energy stored in the charger, and eventually reaches a steady state, as shown in Fig.~\ref{energy_a_b.eps}(a). Figure~\ref{energy_a_b.eps}(c) shows that when the DOPA is introduced ($g=0.1\omega$, $\theta=0$), both the battery energy $E_B^{\text{nr}}$ and the charger energy $E_A^{\text{nr}}$ are enhanced, indicating that the DOPA can increase the stored energy of the quantum battery through the parametric-amplification mechanism.  Figure~\ref{energy_a_b.eps}(e) further illustrates the influence of the DOPA pump phase $\theta$ on the stored energy. For $g=0.1\omega$, as $\theta$ is varied from $0$ to $3\pi/2$, both steady-state energies $E_B^{\text{nr}}(\infty)$ and $E_A^{\text{nr}}(\infty)$ decrease. This behavior is consistent with the preceding analysis of the phase dependence of the steady-state battery energy. Figures~\ref{energy_a_b.eps}(b), \ref{energy_a_b.eps}(d), and \ref{energy_a_b.eps}(f) show the corresponding time evolution of $\eta_{AB}^{\text{nr}}(t)$. Comparing Fig.~\ref{energy_a_b.eps}(d) with Fig.~\ref{energy_a_b.eps}(b), one finds that the introduction of the DOPA slightly reduces $S_d$ from $3.46$ to $3.44$, implying a small decrease in the relative amount of energy stored in the battery. Furthermore, for $g=0.1\omega$, increasing the phase from $\theta=0$ to $\theta=3\pi/2$ reduces $S_d$ from $3.44$ to $3.32$, indicating a further decrease in the battery-to-charger energy ratio.
To elucidate how the DOPA parameters regulate the nonreciprocal charging performance of the quantum battery coupled to a single common reservoir, Fig.~\ref{energy_a_b_theta_g.eps} examines the effects of the nonlinear gain $g$ and the pump-field phase $\theta$ on the stored energy. As shown in Fig.~\ref{energy_a_b_theta_g.eps}(a), when $g$ is increased from $0$ to $0.06\omega$, both the charger energy $E_A^{\text{nr}}$ and the battery energy $E_B^{\text{nr}}$ are enhanced, demonstrating that the parametric-amplification effect of the DOPA becomes stronger with increasing nonlinear gain $g$. Figure~\ref{energy_a_b_theta_g.eps}(b) shows the phase dependence of the stored energy of the nonreciprocal quantum battery coupled to a single common reservoir. For $g=0.02\omega$, the battery energy is enhanced at $\theta=\pi/2$, remains unchanged at $\theta=\pi$, and is reduced at $\theta=3\pi/2$. In addition, as shown in Fig.~\ref{energy_a_b_theta_g.eps}(c), the steady-state ratio $\eta_{AB}^{\text{nr}}(\infty)$ decreases with increasing $g$. By contrast, Fig.~\ref{energy_a_b_theta_g.eps}(d) shows that, within the parameter range considered here, $\eta_{AB}^{\text{nr}}(\infty)$ is nearly insensitive to variations in the pump phase $\theta$.
In addition to the energy storage capacity of the battery, the charging power is also an important figure of merit for evaluating the performance of a quantum battery. A high stored energy does not necessarily imply a sufficiently fast charging process. It is therefore necessary to further examine the average charging power. During the charging process, the average charging power of the quantum battery is defined as
\begin{equation}
    P_B(t)=\frac{E_B(t)}{t},
    \label{eq:avg_power}
\end{equation}
where $E_B(t)$ is the energy stored in the battery at time $t$. By comparing the average charging power under different parameter settings, one can identify the effects of optical parametric amplification, pump-phase control, and optimization of the nonreciprocal dissipative channel on the charging speed of the battery.  Figure~\ref{thesis3/Tex/p_nr.png} shows the time evolution of the average charging power $P_{B}^{\text{nr}}(t)$ of the nonreciprocal quantum battery coupled to a single common reservoir. When the DOPA is introduced ($g=0.1\omega$, $\theta=0$), the parametric-amplification process enhances the coherent energy input at the charger, allowing the battery to acquire more energy within a shorter time and thereby enhancing the average charging power over a certain time interval. Furthermore, for a fixed nonlinear gain $g=0.1\omega$, varying the pump-field phase from $\theta=0$ to $\theta=3\pi/2$ leads to a reduction in the average charging power. The time evolution of the average charging power of the quantum battery is shown for both the reciprocal and nonreciprocal cases. It can be seen that the DOPA further enhances the average charging power of the nonreciprocal quantum battery.

\subsection{Quantum battery performance under reciprocal conditions}
\begin{figure}[ht]
\centerline{
\includegraphics[width=0.46\textwidth]{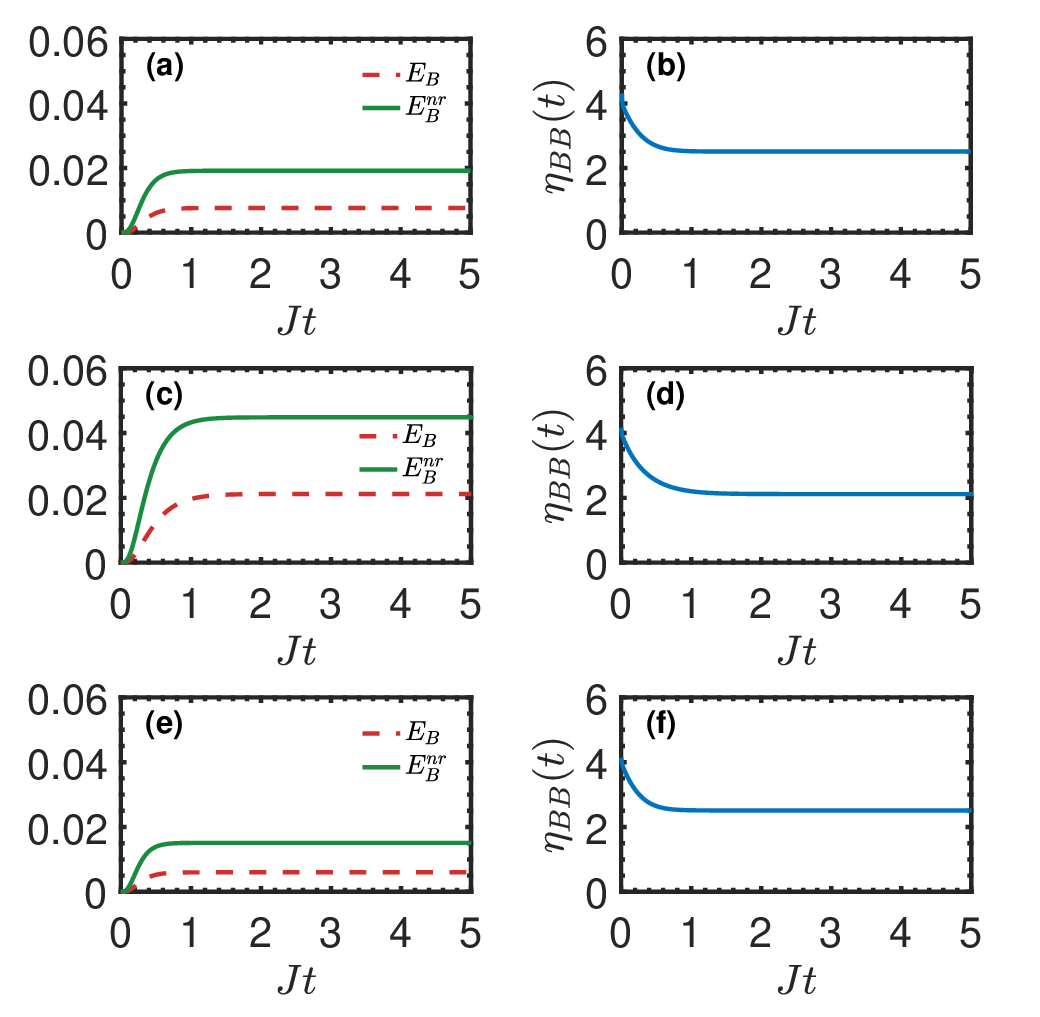}}
\caption{Dynamics of the reciprocal battery energy $E_B(t)$ and 
    the nonreciprocal battery energy $E_B^{\text{nr}}(t)$ in 
    (a)(c)(e), together with the corresponding energy ratio 
    $\eta_{BB}(t) = E_B^{\text{nr}}(t)/E_B(t)$ in (b)(d)(f), 
    as functions of the scaled time $Jt$. 
    The DOPA nonlinear gain $g$ and pump-field phase $\theta$ are 
    set to: (a), (b) $g = 0$, $\theta = 0$; (c), (d) 
    $g = 0.04\omega$, $\theta = 0$; and (e), (f) $g = 0.04\omega$, 
    $\theta = 3\pi/2$. 
    The remaining parameters are $\kappa_a = \kappa_b = 0.3\omega$, $\Gamma_1 = \Gamma_2 = 0.04\omega$, $f = 0.01\omega$, $\phi=0$, $|J| = 0.02\omega$.} \label{energy_b_b.eps}
\end{figure}

\begin{figure}[h]
\centerline{
\includegraphics[width=0.46\textwidth]{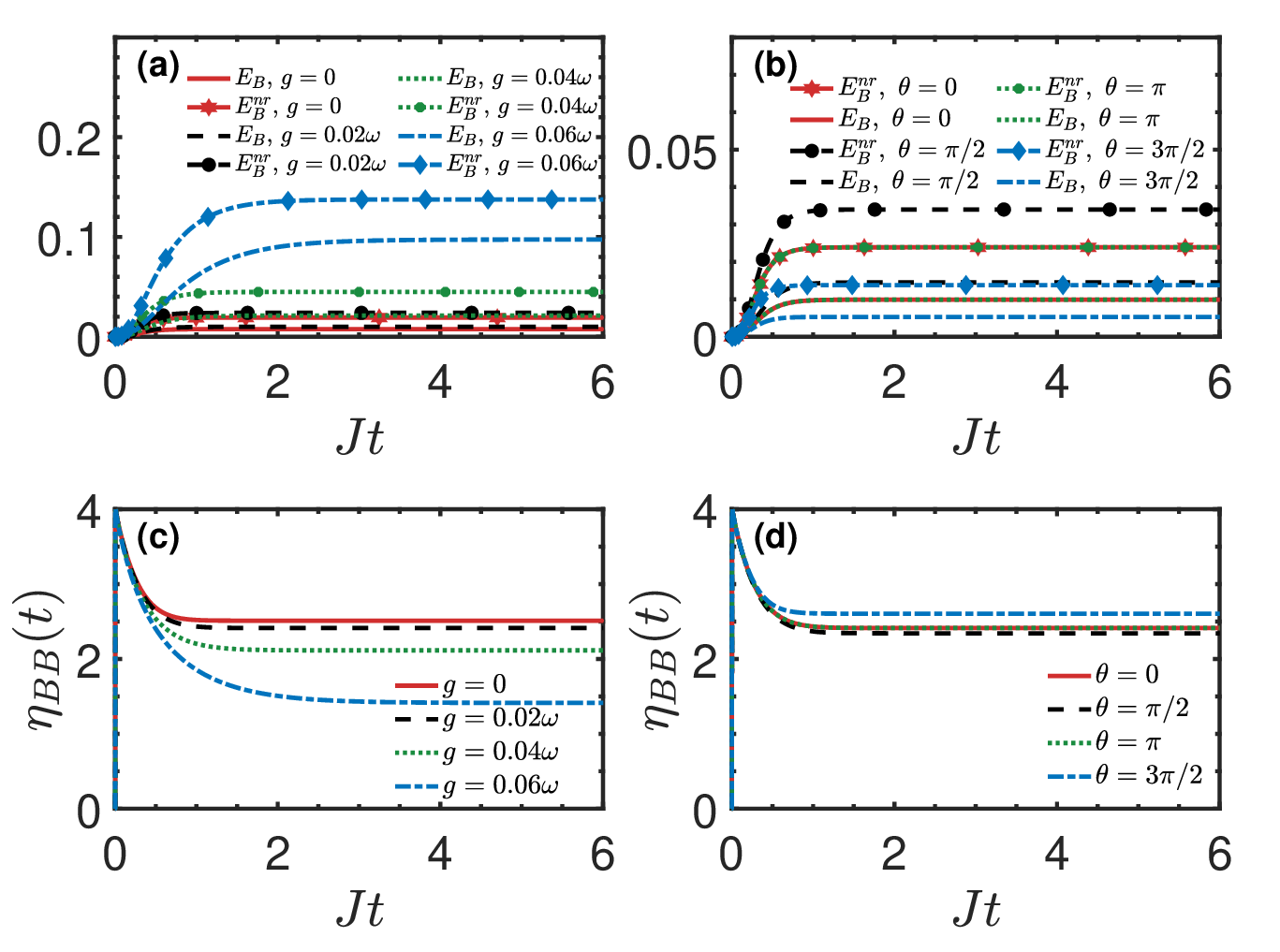}}
\caption{Dynamics of the reciprocal battery energy $E_B(t)$, 
    the nonreciprocal battery energy $E_B^{\text{nr}}(t)$, and 
    the ratio $\eta_{BB}(t) = E_B^{\text{nr}}(t)/E_B(t)$ as 
    functions of the scaled time $Jt$. 
    (a) and (c) show results for different values of the DOPA 
    nonlinear gain $g$ at a fixed pump-field phase $\theta = 0$, 
    while (b) and (d) show results for different values of the 
    pump-field phase $\theta$ at a fixed gain $g = 0.02\omega$. The other parameters are identical to those in Fig.~\ref{energy_b_b.eps}.}\label{energy_b_b_theta_g.eps}
\end{figure}

To evaluate the combined advantage of the nonreciprocal
mechanism and the DOPA, we compare the present system with its
reciprocal counterpart~\cite{Farina035421}.
In the reciprocal case, setting $\Gamma_i = \Gamma = 0$ in
Eq.~\eqref{twohaishenbao} to remove the common reservoir, we
obtain the steady-state battery energy as

\begin{equation}\label{wenenergb}
\begin{aligned}
E_B(\infty)
=&\frac{16\omega|J|^2}{D^2}
[|F|^2 C+4ik_b A Q
]
-\frac{32\omega|J|^2|G|^2 B}{DK},
\end{aligned}
\end{equation}
where $A=4|J|^2+\kappa_a \kappa_b,
B=4|J|^2+2\kappa_a \kappa_b+\kappa_{b}^{2},
C=A^2+16|G|^2\kappa_{b}^{2},
D=A^2-16|G|^2\kappa_{b}^{2},
K=16|G|^2-(\kappa_a+\kappa_b)^2,
Q=(F^*)^2G-F^2G^* $.
Figures~\ref{energy_b_b.eps}(a), \ref{energy_b_b.eps}(c), and \ref{energy_b_b.eps}(e) compare the time evolution of the energy stored in the quantum battery coupled to a single common reservoir under reciprocal and nonreciprocal conditions. As shown in Figs.~\ref{energy_b_b.eps}(a) and \ref{energy_b_b.eps}(c), increasing the nonlinear gain of the DOPA from $g=0$ to $g=0.04\omega$ enhances the steady-state battery energy in both the reciprocal and nonreciprocal cases. Figures~\ref{energy_b_b.eps}(c) and \ref{energy_b_b.eps}(e) further show that, for a fixed $g$, increasing the pump-field phase $\theta$ from $0$ to $3\pi/2$ leads to a reduction in the battery energy in both cases. Moreover, throughout the parameter regimes considered, the energy stored in the battery under the nonreciprocal condition remains higher than that in the reciprocal case.
To quantify the performance advantage of the nonreciprocal
system over its reciprocal counterpart, we define the ratio
$\eta_{BB}(t) = E_B^{\text{nr}}(t)/E_B(t)$, whose
steady-state value is given by
\begin{small}
\begin{align}
\eta_{BB}(\infty)
={}&\frac{6P_1^2P_2^2P_4g^2\Lambda^2(\Lambda^2-16g^2)}{\Lambda^2(\Lambda^2-4g^2)(\Lambda^2-16g^2)^2p}\notag\\
&+\frac{4P_1^2P_2^2P_4f^2(4g^2-\Lambda^2)(\Lambda^2+16g^2)}{\Lambda^2(\Lambda^2-4g^2)(\Lambda^2-16g^2)^2p}\notag\\
&+\frac{32P_1^2P_2^2P_4f^2g\Lambda(4g^2-\Lambda^2)\sin\theta'}{\Lambda^2(\Lambda^2-4g^2)(\Lambda^2-16g^2)^2p},\label{jie11}
\end{align}
\end{small}where $P_1 = 4|J|^2 + (\kappa_a - 4g)\kappa_b, P_2 = 4|J|^2 + (\kappa_a + 4g)\kappa_b,
P_3 = 4|J|^2 + \kappa_b(2\kappa_a + \kappa_b), 
P_4   = 16g^2 - (\kappa_a + \kappa_b)^2, 
q   = 16|J|^4 + 8|J|^2\kappa_a\kappa_b + (16g^2 + \kappa_a^2)\kappa_b^2 + 8g\kappa_b(4|J|^2 + \kappa_a\kappa_b)\sin\theta',p=2g^2P_1P_2P_3-f^2P_4q$, As shown in Figs.~\ref{energy_b_b.eps}(b), \ref{energy_b_b.eps}(d), and \ref{energy_b_b.eps}(f), $\eta_{BB}(t)$ eventually approaches a steady-state value. A comparison between Figs.~\ref{energy_b_b.eps}(b) and \ref{energy_b_b.eps}(d) shows that $\eta_{BB}(\infty)$ decreases from $2.512$ in the absence of the DOPA to $2.115$ at $g=0.04\omega$. This indicates that, within the parameter range considered, increasing the nonlinear gain $g$ reduces the performance advantage of the nonreciprocal quantum battery over its reciprocal counterpart. For a fixed nonlinear gain $g=0.04\omega$, $\eta_{BB}(\infty)$ at $\theta=3\pi/2$ is higher than that at $\theta=0$, demonstrating that the relative performance advantage of the nonreciprocal battery over the reciprocal one can be controlled by tuning the pump-field phase.
Figure~\ref{energy_b_b_theta_g.eps} further provides a systematic comparison of the DOPA-induced modulation of the energy stored in a quantum battery coupled to a single common reservoir under reciprocal and nonreciprocal conditions. As shown in Fig.~\ref{energy_b_b_theta_g.eps}(a), as the nonlinear gain $g$ increases from $0$ to $0.06\omega$, both $E_B^{\text{nr}}(t)$ and $E_B(t)$ gradually increase. Figure~\ref{energy_b_b_theta_g.eps}(c) shows that $\eta_{BB}(t)$ decreases with increasing $g$. The phase dependence shown in Fig.~\ref{energy_b_b_theta_g.eps}(b) is consistent with the preceding analysis: the battery energy is enhanced at $\theta=\pi/2$, remains unchanged at $\theta=\pi$, and is reduced at $\theta=3\pi/2$. As shown in Fig.~\ref{energy_b_b_theta_g.eps}(d), $\eta_{BB}(\infty)$ is enhanced at $\theta=\pi/2$, remains unchanged at $\theta=\pi$, and decreases at $\theta=3\pi/2$. These results demonstrate that tuning the DOPA pump phase provides an effective means of controlling, and under appropriate conditions enhancing, the performance advantage of the nonreciprocal quantum battery over its reciprocal counterpart.

\begin{figure}[h]
\centering
\includegraphics[width=0.4\textwidth]{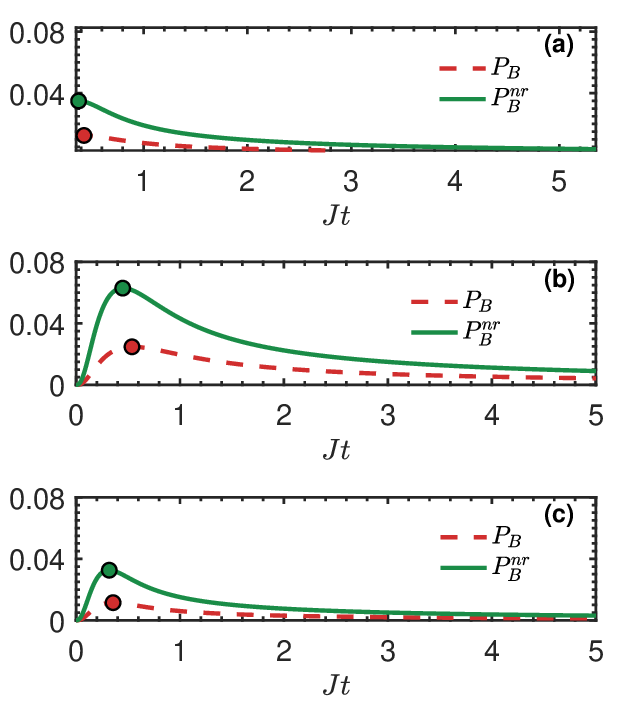}
\caption{Average charging powers $P_B^{\text{nr}}$ in the nonreciprocal case and $P_B$ in the reciprocal case as a function of the scaled time $Jt$, where the nonlinear gain and pump-field phase of the DOPA 
are fixed as (a) $g = 0$, $\theta = 0$; (b) $g = 0.04\omega$, 
$\theta = 0$; (c) $g = 0.04\omega$, $\theta = 3\pi/2$. The other parameters are the same as in Fig.~\ref{energy_b_b.eps}.}
\label{p_nr_non.png}
\end{figure}

Figure~\ref{p_nr_non.png} compares the time evolution of the average charging power of the quantum battery coupled to a single common reservoir under reciprocal and nonreciprocal conditions. Within the parameter range considered, increasing the nonlinear gain $g$ enhances the average charging power in both cases. For the same nonlinear gain $g=0.04\omega$, increasing the pump-field phase from $\theta=0$ to $\theta=3\pi/2$ reduces the average charging power under both reciprocal and nonreciprocal conditions. Moreover, the average charging power of the nonreciprocal quantum battery remains higher than that of its reciprocal counterpart.

\subsection{Optimization of the dissipative-channel asymmetry parameters}
\begin{figure}[h]
\centerline{
\includegraphics[width=0.47\textwidth]{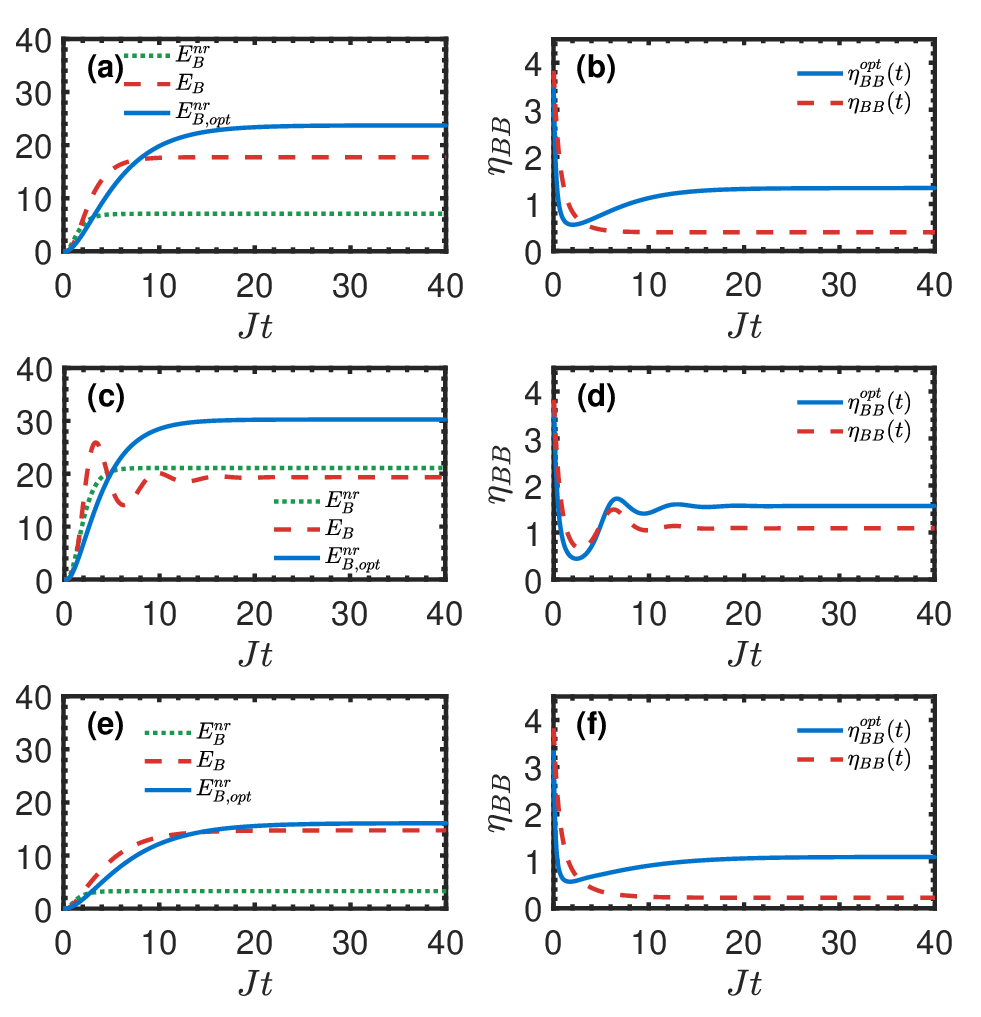}}
\caption{Dynamics of the reciprocal battery energy $E_B(t)$, 
    the nonreciprocal battery energy $E_B^{\text{nr}}(t)$, and 
    the optimal nonreciprocal battery energy 
    $E_{B,\text{opt}}^{\text{nr}}(t)$ in (a)(c)(e), together with 
    the corresponding energy ratios $\eta_{BB}(t)$ and 
    $\eta_{BB}^{\text{opt}}(t)$ in (b)(d)(f), as functions of the 
    scaled time $Jt$. 
    The DOPA nonlinear gain $g$, pump-field phase $\theta$, and 
    asymmetry parameter $\xi$ are set to: (a), (b) $g = 0$, 
    $\theta = 0$, $\xi = 0.173$; (c), (d) $g = 0.02\omega$, 
    $\theta = 0$, $\xi = 0.285$; and (e), (f) $g = 0.02\omega$, 
    $\theta = 3\pi/2$, $\xi = 0.130$. 
    The remaining parameters are ${\kappa_a}= 0.1{\omega}$, ${\kappa_b} = 0.003{\omega}$, ${\Gamma} =0.04{\omega}$, ${\Gamma _1}={\Gamma}/{\xi}$, ${\Gamma _2}={\xi}{\Gamma}$, $f = 0.01\omega$, $\phi=0$, and $|J| = 0.02\omega$.}\label{energy_b_b_optb.eps}
\end{figure}

To further enhance the energy storage of the quantum battery,
we introduce an optimization parameter $\xi > 0$ by setting
$\Gamma_1 = \Gamma/\xi$ and $\Gamma_2 = \Gamma\xi$.
This parametrization keeps the product
$\Gamma_1\Gamma_2 = \Gamma^2$ constant while allowing the
asymmetry between the two dissipative channels to be
continuously varied through $\xi$.
Substituting $\Gamma_1 = \Gamma/\xi$ and $\Gamma_2 = \Gamma\xi$
into Eq.~\eqref{wenenergyb} and setting the partial derivative
with respect to $\xi$ to zero, we obtain the optimal
condition
\begin{equation}
    \frac{\partial E_B^{\text{nr}}(\infty)}{\partial \xi}\bigg|_{\xi = \xi_{\text{opt}}} = 0.
    \label{eq:opt_condition12}
\end{equation}
Since Eq.~\eqref{eq:opt_condition12} does not admit a
closed-form solution, we numerically solve for the optimal
parameter $\xi_{\text{opt}}$ and thereby determine the optimal
dissipation configuration
$(\Gamma_1^{\text{opt}},\,\Gamma_2^{\text{opt}})$.
Substituting the resulting $\xi_{\text{opt}}$ into
Eqs.~\eqref{second1} and~\eqref{Ei}, we obtain the optimal
nonreciprocal battery energy $E_{B,\text{opt}}^{\text{nr}}(t)$.
As shown in Fig.~\ref{energy_b_b_optb.eps},
$E_{B,\text{opt}}^{\text{nr}}(t)$ exceeds the unoptimized
value $E_B^{\text{nr}}(t)$, demonstrating that appropriately
tuning the asymmetry between the two dissipative channels
provides an effective means of enhancing the quantum battery
energy storage.
Figures~\ref{energy_b_b_optb.eps}(a), \ref{energy_b_b_optb.eps}(c), and \ref{energy_b_b_optb.eps}(e) show the time evolution of the battery energy for three different configurations of a quantum battery coupled to a single common reservoir: the reciprocal battery energy $E_B(t)$, the nonreciprocal battery energy $E_B^{\text{nr}}(t)$, and the optimized nonreciprocal battery energy $E_{B,\text{opt}}^{\text{nr}}(t)$.   Figures~\ref{energy_b_b_optb.eps}(b), \ref{energy_b_b_optb.eps}(d), and \ref{energy_b_b_optb.eps}(f) show the corresponding time evolution of the energy ratios
$\eta_{BB}^{\text{opt}}=E_{B,\text{opt}}^{\text{nr}}(t)/E_B(t)$
and
$\eta_{BB}=E_B^{\text{nr}}(t)/E_B(t)$.
The results demonstrate that the optimized nonreciprocal battery consistently exhibits a higher energy-storage capacity than the other two configurations. As shown in Fig.~\ref{energy_b_b_optb.eps}(a), even in the absence of the DOPA, parameter optimization enables the nonreciprocal battery to achieve the highest steady-state stored energy. By contrast, for the parameters considered here, the unoptimized nonreciprocal battery stores less energy than the reciprocal battery, highlighting the crucial role of parameter optimization in exploiting the advantages of nonreciprocal energy transfer. When the DOPA is introduced with $g=0.02\omega$, Fig.~\ref{energy_b_b_optb.eps}(c) shows that the steady-state stored energies of all three configurations are enhanced, while the optimized nonreciprocal battery continues to exhibit the highest stored energy. A comparison between Figs.~\ref{energy_b_b_optb.eps}(c) and \ref{energy_b_b_optb.eps}(e) further shows that, for $g=0.02\omega$, varying the DOPA pump phase from $\theta=0$ to $\theta=3\pi/2$ reduces the battery energy in all three configurations. Figures~\ref{energy_b_b_optb.eps}(b), \ref{energy_b_b_optb.eps}(d), and \ref{energy_b_b_optb.eps}(f) further illustrate the behavior of the ratios $\eta_{BB}^{\text{opt}}$ and $\eta_{BB}$. Comparing Fig.~\ref{energy_b_b_optb.eps}(d) with Fig.~\ref{energy_b_b_optb.eps}(b), one finds that increasing the nonlinear gain from $g=0$ to $g=0.02\omega$ enhances the steady-state values of both ratios. In contrast, a comparison between Figs.~\ref{energy_b_b_optb.eps}(f) and \ref{energy_b_b_optb.eps}(d) shows that varying the pump phase from $\theta=0$ to $\theta=3\pi/2$ decreases both ratios. These results demonstrate that the nonlinear gain $g$ and pump phase $\theta$ of the DOPA provide flexible control parameters for regulating the energy-storage performance of the quantum battery. When combined with parameter optimization, the DOPA provides an additional degree of control for further exploiting the energy-storage potential of the nonreciprocal quantum battery.
\section{DOPA Effects in Dual Common Reservoir Configuration}
\subsection{Two-common-reservoir model and analytical steady-state solution}
The spectral response functions are taken as $
g(\omega') = \sqrt{{\Gamma_1}/{2\pi}},
 G(\omega') = \sqrt{{\Gamma_2}/{2\pi}},
 v(\omega') = \sqrt{{\Gamma_3}/{2\pi}},
 u(\omega') = ({-2iJ-\sqrt{\Gamma_1\Gamma_2}})/{\sqrt{2\pi\Gamma_3}}.$ The spectral densities of the reservoirs are $J_1(\omega' ) = {{{\Gamma _1}}}/{{2\pi }},
J_2(\omega' ) = {{{\Gamma _2}}}/{{2\pi }},
J_3(\omega' ) ={{{\Gamma _3}}}/{{2\pi }},
J_4(\omega' )={2JJ^*}/{\pi\Gamma_3} + {iJ\sqrt{\Gamma_1\Gamma_2}}/{\pi\Gamma_3} - {iJ^*\sqrt{\Gamma_1\Gamma_2}}/{\pi\Gamma_3} + {\Gamma_1\Gamma_2}/{2\pi\Gamma_3},
J_5(\omega') = J_6(\omega') = {\sqrt{\Gamma_1 \Gamma_2}}/{2\pi},
J_7(\omega' )=-{iJ}/{\pi} - {\sqrt{\Gamma_1\Gamma_2}}/{2\pi},
J_8(\omega' )={iJ^*}/{\pi} - {\sqrt{\Gamma_1\Gamma_2}}/{2\pi}.$
The dynamics of the nonreciprocal system is governed by equations of motion for the first- and second-order moments with
\begin{small}
\begin{align}
\frac{d\langle\hat{a}\rangle}{dt}=&-(i\Delta+\frac{\Gamma_1}{2}+\frac{\Gamma_3}{2}+\frac{\kappa_a}{2})\langle\hat{a}\rangle-iF-2iG\langle\hat{a}^\dagger\rangle,\nonumber\\
\frac{d\langle\hat{b}\rangle}{dt}=&-(i\Delta+\frac{\Gamma_2}{2}+\frac{\Gamma_4}{2}+\frac{\kappa_b}{2})\langle\hat{b}\rangle-2iJ^*\langle\hat{a}\rangle,\nonumber\\
\frac{d\langle\hat{a}^\dagger\hat{a}\rangle}{dt}=&-(\Gamma_1+\Gamma_3+\kappa_a)\langle\hat{a}^\dagger\hat{a}\rangle-iF\langle\hat{a}^\dagger\rangle+iF^*\langle\hat{a}\rangle\nonumber\\
&-2iG\langle\hat{a}^{\dagger 2}\rangle+2iG^*\langle\hat{a}^2\rangle,\nonumber\\
\frac{d\langle\hat{b}^\dagger\hat{b}\rangle}{dt}=&-(\Gamma_2+\Gamma_4+\kappa_b)\langle\hat{b}^\dagger\hat{b}\rangle-2iJ^*\langle\hat{b}^\dagger\hat{a}\rangle+2iJ\langle\hat{a}^\dagger\hat{b}\rangle,\nonumber\\
\frac{d\langle\hat{a}^\dagger\hat{b}\rangle}{dt}=&-(\frac{\Gamma_1}{2}+\frac{\Gamma_2}{2}+\frac{\Gamma_3}{2}+\frac{\Gamma_4}{2}+\frac{\kappa_a}{2}+\frac{\kappa_b}{2})\langle\hat{a}^\dagger\hat{b}\rangle\nonumber\\
&-2iJ^*\langle\hat{a}^\dagger\hat{a}\rangle+iF^*\langle\hat{b}\rangle+2iG^*\langle\hat{a}\hat{b}\rangle,\nonumber\\
\frac{d\langle\hat{a}^2\rangle}{dt}=&-(2i\Delta+\Gamma_1+\Gamma_3+\kappa_a)\langle\hat{a}^2\rangle-2iF\langle\hat{a}\rangle\nonumber\\
&-4iG\langle\hat{a}^\dagger\hat{a}\rangle-2iG,\nonumber\\
\frac{d\langle\hat{a}\hat{b}\rangle}{dt}=&-(2i\Delta+\frac{\Gamma_1}{2}+\frac{\Gamma_2}{2}+\frac{\Gamma_3}{2}+\frac{\Gamma_4}{2}+\frac{\kappa_a}{2}+\frac{\kappa_b}{2})\langle\hat{a}\hat{b}\rangle\nonumber\\
&-2iJ^*\langle\hat{a}^2\rangle-iF\langle\hat{b}\rangle-2iG\langle\hat{a}^\dagger\hat{b}\rangle,
\label{second12}
\end{align}
\end{small}where $\Gamma_4 = {(2J-i\sqrt{\Gamma_1\Gamma_2}\bigr)(2J^*+i\sqrt{\Gamma_1\Gamma_2}\bigr)}/{\Gamma_3}$, $\Lambda_a = \Gamma_1 + \Gamma_3 + \kappa_a$ and
$\Lambda_b = \Gamma_2 + \Gamma_4 + \kappa_b$ denote the effective
dissipation rates of the charger and battery, respectively. As in the single-common-reservoir case, the evolution equation
for the charger contains no battery variables, and the
nonreciprocal transmission structure is preserved.
Solving Eq.~\eqref{second12} for the steady state, we obtain
the charger and battery energies as
\begin{align}
 E_B^{\mathrm{nr}}(\infty) =&\frac{128 \omega |J|^2|G|^2(2\Lambda_a+\Lambda_b)}
{\Lambda_b(16|G|^2-\Lambda_a^2)
[16|G|^2-(\Lambda_a+\Lambda_b)^2]}\nonumber\\
&+\frac{64 \omega |J|^2 |F|^2(16|G|^2+\Lambda_a^2)}{(16|G|^2-\Lambda_a^2)^2\Lambda_b^2}\nonumber\\
&+\frac{256 i \omega |J|^2 \Lambda_a
[(F^*)^2G-F^2G^*]}
{(16|G|^2-\Lambda_a^2)^2\Lambda_b^2},
  \label{eq:EB_two_res}\\[6pt]
E_A^{\mathrm{nr}}(\infty)={}&\frac{64\omega |G|^2(|F|^2-2|G|^2)}{(-16|G|^2+\Lambda_a^2)^2}\notag\\
&+\frac{16i\omega[(F^*)^2G-F^2G^*]\Lambda_a}{(-16|G|^2+\Lambda_a^2)^2}\notag\\
&+\frac{4\omega(|F|^2+2|G|^2)\Lambda_a^2}{(-16|G|^2+\Lambda_a^2)^2}.
  \label{eq:EA_two_res}
\end{align}
Comparing Eq.~\eqref{eq:EA_two_res} with the single-reservoir
result in Eq.~\eqref{wenenergya}, $E_A^{\mathrm{nr}}(\infty)$
remains independent of all battery parameters ($\kappa_b$,
$\Gamma_2$, and so forth), confirming that the steady-state
behavior of the charger is determined solely by its own
parameters---a hallmark of the nonreciprocal dissipative
structure. 
We obtain the steady-state battery energy as
\begin{equation}
\begin{aligned}
 E_B^{\mathrm{nr}}(\infty)=&\frac{128\omega|J|^2  g^2  (2\Lambda_a + \Lambda_b)}{(16 g^2 - \Lambda_a^2) \Lambda_b 
[16 g^2 - (\Lambda_a + \Lambda_b)^2]}\\
&+\frac{64\omega |J|^2 f^2(16g^2 + \Lambda_a^2 + 8g\Lambda_a\sin\theta')}{(-16 g^2+\Lambda_a^2)^2 \Lambda_b^2 }.
\end{aligned}
\label{eq:EB_two_expand}
\end{equation}
Specifically, when $\sin\theta' = 1$ (e.g., $\theta = \pi/2$, $\phi = 0$), this interference term reaches its maximum positive value, and parametric amplification interferes constructively with the driving field, yielding peak battery energy storage. Conversely, when $\sin\theta' = 0$ ($\theta' = 0$ or $\pi$), the interference term vanishes, and the stored energy is governed solely by the phase-independent terms. Finally, when $\sin\theta' = -1$ ($\theta' = 3\pi/2$), the interference becomes maximally negative, producing a destructive effect that minimizes the stored energy. In this destructive limit, the numerator reduces to the perfect square $f^2(\Lambda_a - 4g)^2$, ensuring that the stored energy remains strictly positive. 

This phase dependence is in full agreement with the single-reservoir scenario: because the DOPA acts solely on the charger mode, the relative phase between the pump and driving fields fundamentally dictates the nature of their interference---a mechanism strictly independent of the number of common reservoirs.
\begin{figure}[h]
\centerline{
\includegraphics[width=0.46\textwidth]{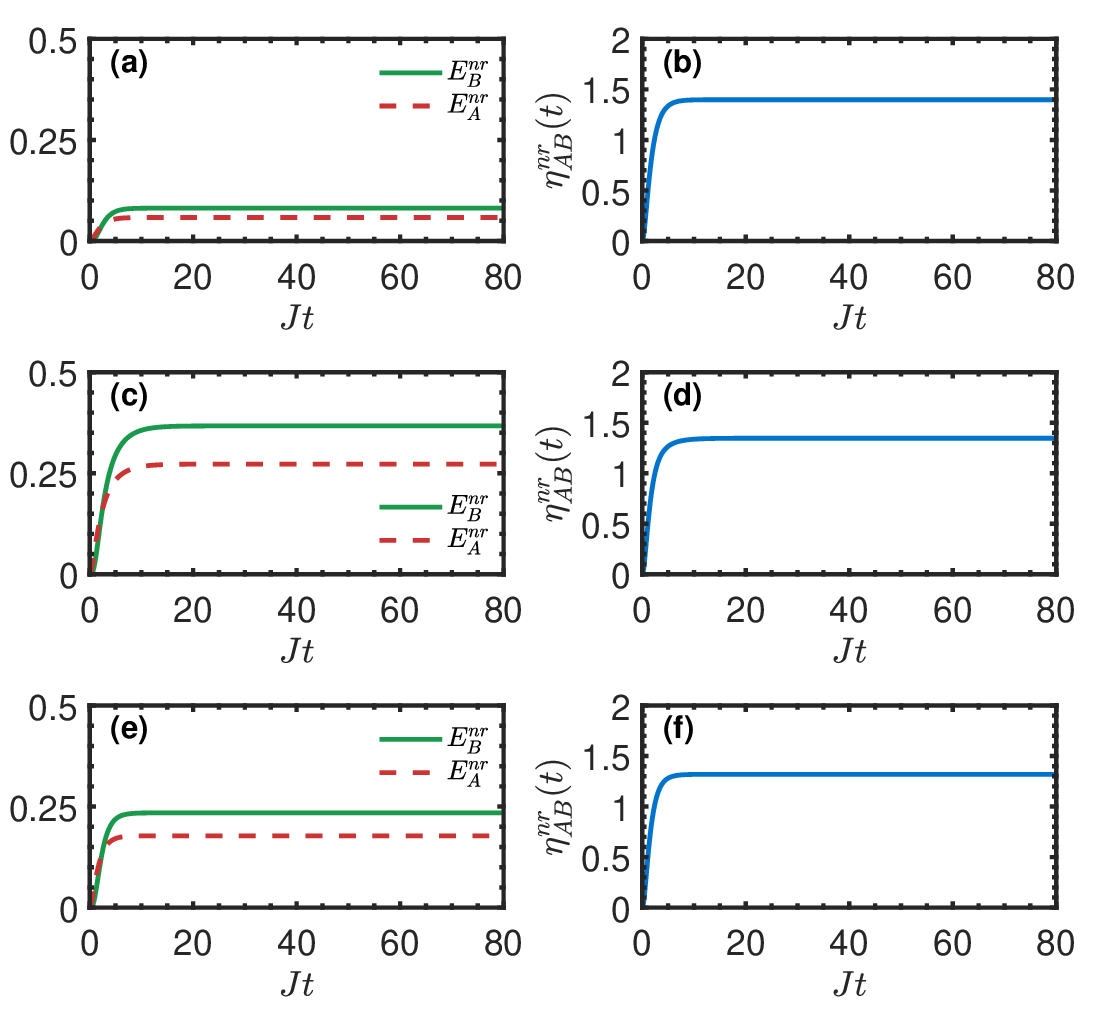}}
\caption{In the nonreciprocal regime, the total energies of the charger 
$E_A^{\text{nr}}$ and battery $E_B^{\text{nr}}$, and their ratio 
$\eta_{AB}^{\text{nr}}(t) = E_B^{\text{nr}}/E_A^{\text{nr}}$ are 
plotted versus the scaled time $Jt$. 
DOPA parameters: (a) and (b) $g = 0$, $\theta = 0$; 
(c) and (d) $g = 0.1\omega$, $\theta = 0$; 
(e) and (f) $g = 0.1\omega$, $\theta = 3\pi/2$. 
The other parameters are ${\kappa_a} ={\kappa_b} = 0.03{\omega}$, ${\Gamma _1} ={\Gamma _2}={\Gamma _3} =0.4{\omega}$, $f = 0.01\omega$, $\phi=0$, and  ${|J|}=0.6\omega$. }\label{Two_energy_a_b.eps}
\end{figure}
\begin{figure}[t]
\centerline{
\includegraphics[width=0.46\textwidth]{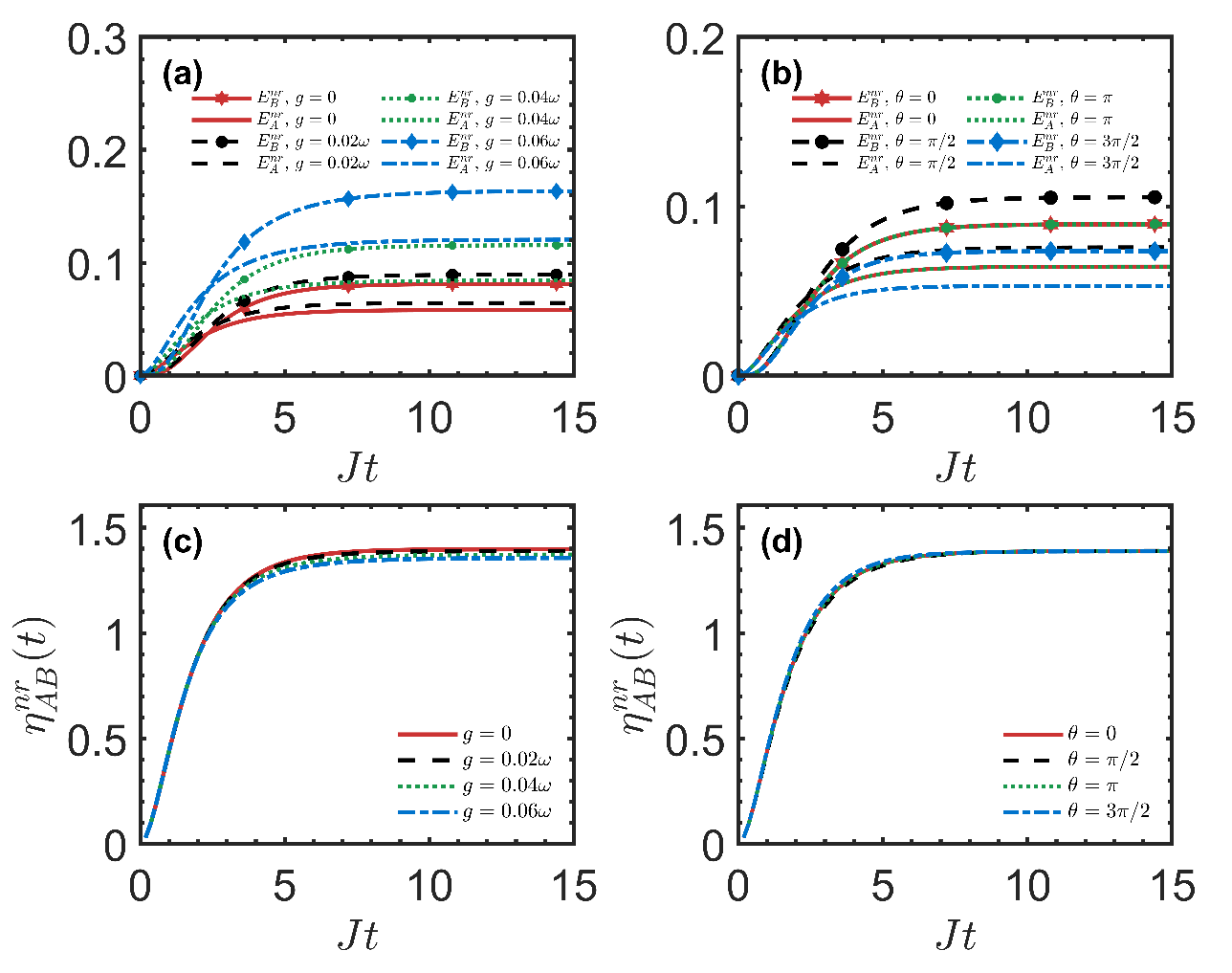}}
\caption{The total energy of the charger $E_A^{\text{nr}}$, the battery 
$E_B^{\text{nr}}$, and the energy ratio 
$\eta_{AB}^{\text{nr}} = E_B^{\text{nr}}/E_A^{\text{nr}}$ are plotted 
against the scaled time $Jt$ in the nonreciprocal regime. 
(a) and (c) show results for varying DOPA nonlinear gain $g$ with 
$\theta = 0$. 
(b) and (d) show results for varying pump-field phase $\theta$ with 
$g = 0.02\omega$. The other parameters are the same as  Fig.~\ref{Two_energy_a_b.eps}.} \label{Two_energy_a_b_theta_G.eps}
\end{figure}We next analyze the energy distribution between the charger
and the battery.
Defining the energy ratio
$\eta_{AB}^{\text{nr}}(t) = E_B^{\text{nr}}(t)/E_A^{\text{nr}}(t)$
and combining Eq.~\eqref{eq:EB_two_res} with
Eq.~\eqref{eq:EA_two_res}, we obtain
\begin{equation}
\eta_{AB}^{\mathrm{nr}}(\infty)
=\frac{16|J|^2}{\Lambda_b^2}-\frac{32|J|^2g^2(\Lambda_a^2-16g^2)^2}{\Lambda_b^2[(\Lambda_a+\Lambda_b)^2-16g^2]\mathcal{Q}_a
},
\label{eq:eta_two_res}
\end{equation}where $\mathcal{Q}_a=
2g^2(\Lambda_a^2-16g^2)
+f^2(16g^2+\Lambda_a^2)
+8f^2g\Lambda_a\sin\theta'$,

\begin{figure}[!htbp]
\centering
\includegraphics[width=0.4\textwidth]{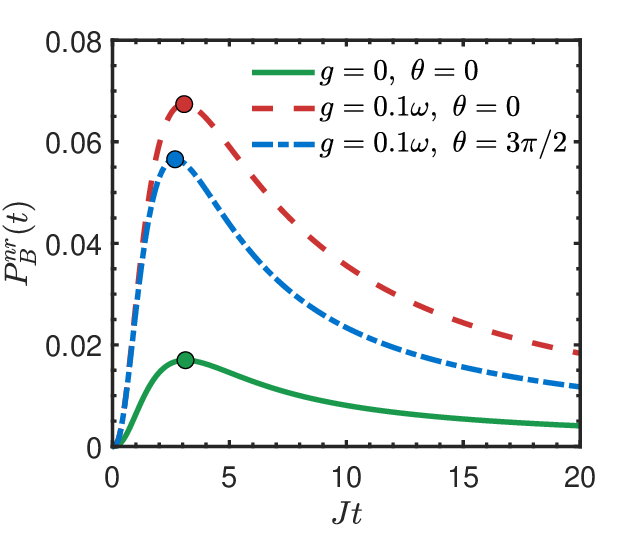}
\caption{Average charging power $P_B^{\text{nr}}$ as a function of the scaled time $Jt$ in the nonreciprocal case. The other parameters are the same as in Fig.~\ref{Two_energy_a_b.eps}.}
\label{Two_p_nr.png}
\end{figure}

The preceding analysis establishes a general principle applicable to both the single- and two-reservoir cases: in a quantum battery system operating under nonreciprocal dissipation, DOPA parametric amplification uniformly enhances the energy stored in both the charger and the battery . This result not only deepens our fundamental understanding of nonreciprocal energy transfer but also provides theoretical guidance for designing efficient and tunable quantum battery systems.


Figure~\ref{Two_energy_a_b.eps} shows the time evolution of the stored energies of the nonreciprocal quantum battery in the two-common-reservoir configuration. Similar to the single-common-reservoir case, the presence of the DOPA enhances the steady-state energy stored in the battery. For a fixed nonlinear gain $g=0.1\omega$, varying the DOPA pump phase from $\theta=0$ to $\theta=3\pi/2$ leads to a reduction in the battery energy. Figures~\ref{Two_energy_a_b.eps}(b), \ref{Two_energy_a_b.eps}(d), and \ref{Two_energy_a_b.eps}(f) show the corresponding time evolution of $\eta_{AB}^{\text{nr}}(t)$. The introduction of the DOPA reduces $\eta_{AB}^{\text{nr}}(t)$. Moreover, for $g=0.1\omega$, increasing the pump phase from $\theta=0$ to $\theta=3\pi/2$ further decreases $\eta_{AB}^{\text{nr}}(t)$. These results demonstrate that the DOPA-induced enhancement of the battery energy persists in the two-common-reservoir configuration, although the relative energy ratio between the battery and the charger is reduced.    Figure~\ref{Two_energy_a_b_theta_G.eps} further illustrates the dependence of the stored energy on the DOPA parameters. As shown in Fig.~\ref{Two_energy_a_b_theta_G.eps}(a), increasing the nonlinear gain $g$ from $0$ to $0.06\omega$ enhances the energies stored in both the charger and the battery. The phase dependence shown in Fig.~\ref{Two_energy_a_b_theta_G.eps}(b) is consistent with that obtained for the single-common-reservoir configuration: the battery energy is enhanced at $\theta=\pi/2$, remains unchanged at $\theta=\pi$, and is reduced at $\theta=3\pi/2$. As shown in Fig.~\ref{Two_energy_a_b_theta_G.eps}(c), the steady-state ratio $\eta_{AB}^{\text{nr}}(\infty)$ increases as the nonlinear gain $g$ decreases. By contrast, Fig.~\ref{Two_energy_a_b_theta_G.eps}(d) shows that, for the parameters considered here, $\eta_{AB}^{\text{nr}}(\infty)$ is essentially independent of the pump phase $\theta$.  Figure~\ref{Two_p_nr.png} shows the time evolution of the average charging power $P_B^{\text{nr}}(t)$ of the nonreciprocal quantum battery coupled to two common reservoirs. Similar to the single-common-reservoir case, the average charging power increases rapidly during the initial charging stage, reaches its maximum within a relatively short time, and subsequently decreases as the system approaches the steady state. This behavior indicates that, once the nonreciprocal energy-transfer channel is established, the battery rapidly absorbs energy from the charger at the early stage of the charging process. As the system approaches the steady state, the rate of energy accumulation in the battery gradually decreases, leading to a reduction in the average charging power. The DOPA provides an additional means of controlling the charging power in the two-common-reservoir configuration. In the presence of the DOPA, parametric amplification at the charger increases the energy available for transfer to the battery, thereby enhancing the average charging power. In contrast, when the DOPA pump phase leads to destructive interference between the parametric-amplification process and the single-photon driving field, the effective energy injected into the charger is reduced, resulting in a lower average charging power. In particular, for $g=0.1\omega$, increasing the pump phase from $\theta=0$ to $\theta=3\pi/2$ reduces the average charging power. These results demonstrate that the DOPA remains an effective means of controlling the charging rate of the nonreciprocal quantum battery in the two-common-reservoir configuration, with a modulation mechanism consistent with that found in the single-common-reservoir case.

\subsection{Quantum battery performance under reciprocal conditions: two-reservoir case}
To quantify the performance advantage of the nonreciprocal
mechanism over the reciprocal case in the two-common-reservoir
configuration, we define the ratio
$\eta_{BB}(t) = E_B^{\text{nr}}(t)/E_B(t)$.
Since the reciprocal battery is obtained by removing all common
reservoirs ($\Gamma_i = 0$), its steady-state energy remains
in Eq.~(\ref{wenenergb}).
\begin{figure}[t]
\centerline{
\includegraphics[width=0.46\textwidth]{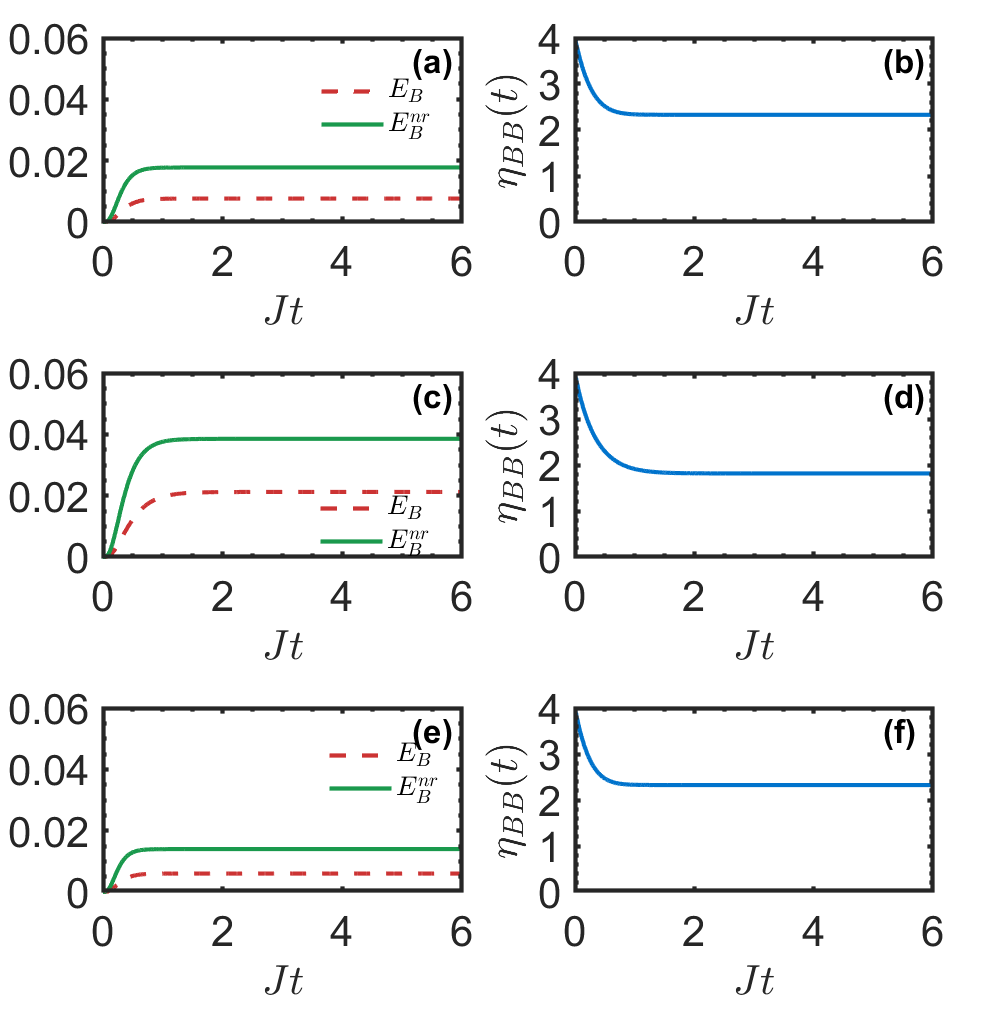}}
\caption{ The battery energy in the reciprocal $E_B(t)$ and 
nonreciprocal $E_B^{\text{nr}}(t)$ regimes plotted versus the scaled 
time $Jt$, where the nonlinear gain and pump-field phase of the DOPA 
are fixed as (a) $g = 0$, $\theta = 0$; (c) $g = 0.04\omega$, 
$\theta = 0$; (e) $g = 0.04\omega$, $\theta = 3\pi/2$.  The ratio $\eta_{BB}(t) = E_B^{\text{nr}}(t)/E_B(t)$ 
versus the scaled time $Jt$, for different parameter values: 
(b) $g = 0$, $\theta = 0$; (d) $g = 0.04\omega$, $\theta = 0$; 
(f) $g = 0.04\omega$, $\theta = 3\pi/2$. 
The other parameters are ${\kappa_a} ={\kappa_b} = 0.3{\omega}$, ${\Gamma _1} ={\Gamma _2}={\Gamma _3} =0.03{\omega}$, $f = 0.01\omega$, $\phi=0$, 
and $|J| = 0.02\omega$.} \label{Two_energy_b_b.eps}
\end{figure}
The steady-state value of $\eta_{BB}$ is given by
\begin{equation}
\begin{aligned}
\eta_{BB}(\infty)={}&-\frac{8g^2P_1^2P_2^2P_4(2\Lambda_a+\Lambda_b)}{X_a\Lambda_b X_{ab}Y}
\\
&+\frac{4f^2P_1^2P_2^2P_4(16g^2+\Lambda_a^2+8g\Lambda_a\sin\theta')}{X_a^2\Lambda_b^2Y},
\end{aligned}
\label{jie12}
\end{equation}
where $P_1 = 4|J|^2 + (\kappa_a - 4g)\kappa_b,P_2 = 4|J|^2 + (\kappa_a + 4g)\kappa_b,
P_4   = 16g^2 - (\kappa_a + \kappa_b)^2,X_a=\Lambda_a^2-16g^2,
X_{ab}=16g^2-(\Lambda_a+\Lambda_b)^2,Y=f^2P_4q-2g^2P_1P_2P_3$.  Figures~\ref{Two_energy_b_b.eps}(a), \ref{Two_energy_b_b.eps}(c), and \ref{Two_energy_b_b.eps}(e) compare the time evolution of the battery energy in the two-common-reservoir configuration under reciprocal and nonreciprocal conditions. A comparison between Figs.~\ref{Two_energy_b_b.eps}(a) and \ref{Two_energy_b_b.eps}(c) shows that the introduction of the DOPA enhances the energy stored in the battery in both the reciprocal and nonreciprocal cases. Comparing Figs.~\ref{Two_energy_b_b.eps}(c) and \ref{Two_energy_b_b.eps}(e), one finds that, for $g=0.04\omega$, increasing the pump phase from $\theta=0$ to $\theta=3\pi/2$ reduces the battery energy in both cases. Figures~\ref{Two_energy_b_b.eps}(b), \ref{Two_energy_b_b.eps}(d), and \ref{Two_energy_b_b.eps}(f) show the corresponding time evolution of $\eta_{BB}(t)$. As $g$ increases from $0$ to $0.04\omega$, the steady-state value $\eta_{BB}(\infty)$ decreases. In contrast, for a fixed nonlinear gain $g=0.04\omega$, varying the pump phase from $\theta=0$ to $\theta=3\pi/2$ increases $\eta_{BB}(\infty)$. These results indicate that the DOPA parameters not only control the energy stored in the battery but also regulate the relative performance advantage of the nonreciprocal mechanism over its reciprocal counterpart.
Figure~\ref{Two_energy_b_b_theta_G.eps} further illustrates the effects of the DOPA parameters on the battery energy under reciprocal and nonreciprocal conditions in the two-common-reservoir configuration. As shown in Fig.~\ref{Two_energy_b_b_theta_G.eps}(a), for $\theta=0$, increasing the nonlinear gain $g$ from $0$ to $0.06\omega$ gradually enhances both $E_B^{\text{nr}}(t)$ and $E_B(t)$. Moreover, the steady-state value of $E_B^{\text{nr}}(t)$ remains higher than that of $E_B(t)$ throughout the parameter range considered. Figure~\ref{Two_energy_b_b_theta_G.eps}(c) shows that $\eta_{BB}(t)$ decreases with increasing $g$. The phase dependence shown in Fig.~\ref{Two_energy_b_b_theta_G.eps}(b) is consistent with that found in the single-common-reservoir configuration: the battery energy is enhanced at $\theta=\pi/2$, remains unchanged at $\theta=\pi$, and is reduced at $\theta=3\pi/2$. As shown in Fig.~\ref{Two_energy_b_b_theta_G.eps}(d), $\eta_{BB}(t)$ can also be controlled by tuning the pump phase $\theta$. For a fixed nonlinear gain $g=0.02\omega$, $\eta_{BB}(t)$ increases at $\theta=3\pi/2$, remains unchanged at $\theta=\pi$, and decreases at $\theta=\pi/2$. These results further demonstrate the capability of the DOPA to control both the absolute energy-storage performance and the relative nonreciprocal enhancement of the quantum battery coupled to two common reservoirs.
\begin{figure}[t]
\centerline{
\includegraphics[width=0.46\textwidth]{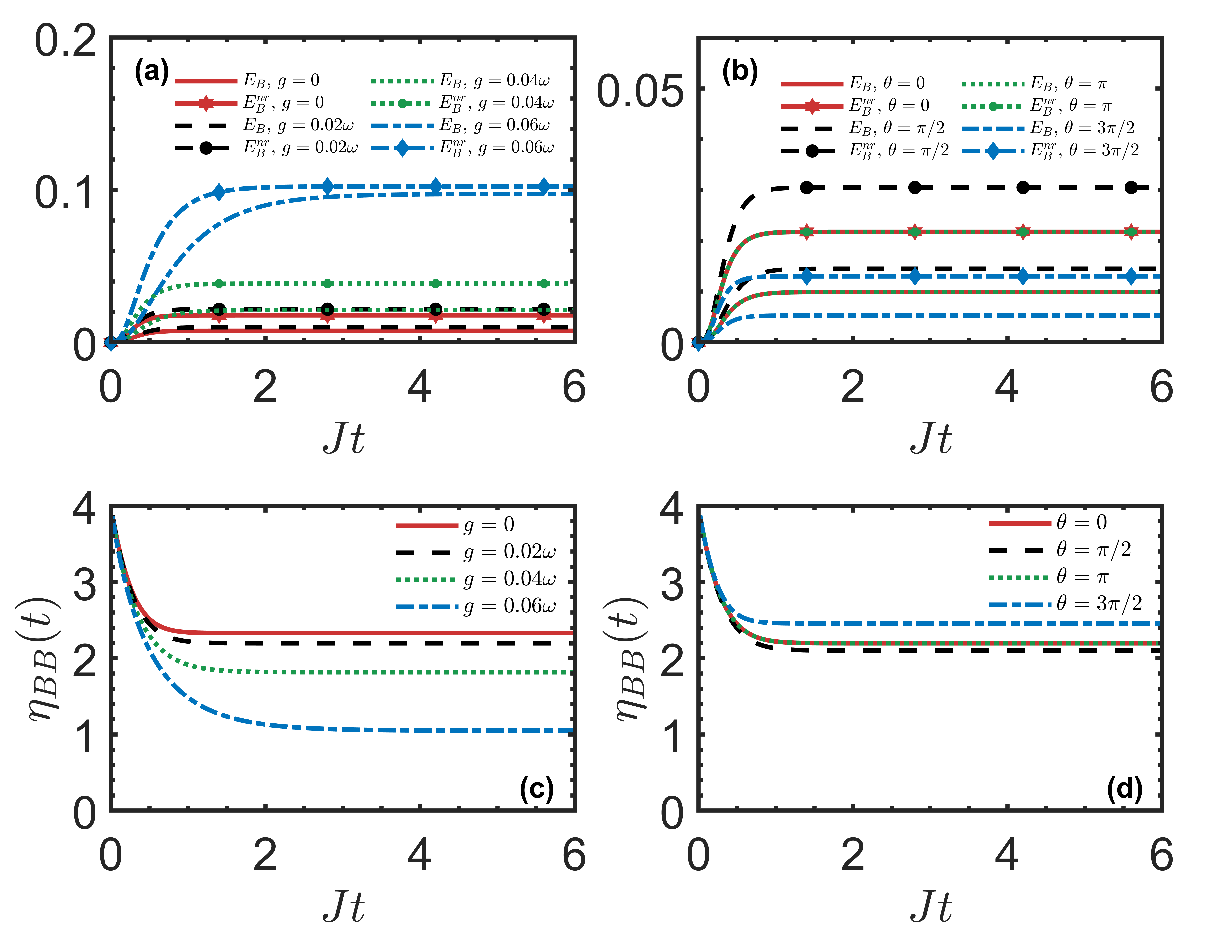}}
\caption{The battery energy in the reciprocal regime $E_B(t)$ and 
nonreciprocal regime $E_B^{\text{nr}}(t)$, along with the ratio 
$\eta_{BB}(t) = E_B^{\text{nr}}(t)/E_B(t)$, are plotted versus the 
scaled time $Jt$. (a) and (b) show results for different DOPA nonlinear 
gain values $g$ with $\theta = 0$, while (c) and (d) show results for 
different DOPA pump-field phase values $\theta$ with $g = 0.02\omega$. The other parameters are the same as Fig.~\ref{Two_energy_b_b.eps}.} \label{Two_energy_b_b_theta_G.eps}
\end{figure}
\begin{figure}[!htbp]
\centering
\includegraphics[width=0.4\textwidth]{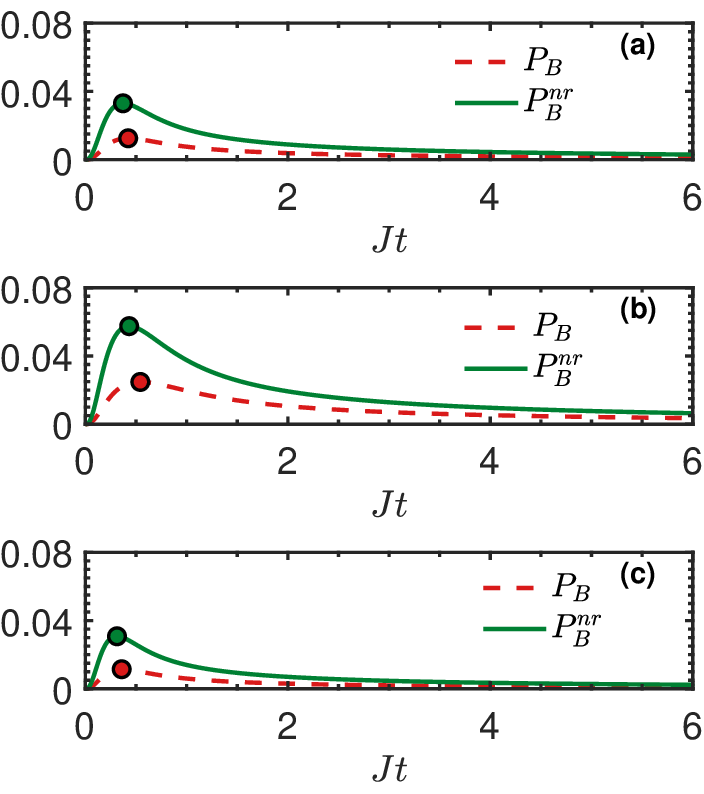}
\caption{Average charging powers $P_B^{\text{nr}}$ in the nonreciprocal case and $P_B$ in the reciprocal case as a function of the scaled time $Jt$, where the nonlinear gain and pump-field phase of the DOPA 
are fixed as (a) $g = 0$, $\theta = 0$; (b) $g = 0.04\omega$, 
$\theta = 0$; (c) $g = 0.04\omega$, $\theta = 3\pi/2$. The other parameters are the same as in Fig.~\ref{Two_energy_b_b.eps}.}
\label{Two_p_nr_non.png}
\end{figure}

We further investigate the time evolution of the average charging power of the quantum battery under reciprocal and nonreciprocal conditions for different DOPA parameters. As shown in Figs.~\ref{Two_p_nr_non.png}(a) and \ref{Two_p_nr_non.png}(b), the introduction of the DOPA enhances the average charging power in both cases, while the average charging power under the nonreciprocal condition remains higher than that under the reciprocal condition. A comparison between Figs.~\ref{Two_p_nr_non.png}(b) and \ref{Two_p_nr_non.png}(c) shows that, for $g=0.04\omega$, increasing the pump phase from $\theta=0$ to $\theta=3\pi/2$ reduces the average charging power in both cases. Nevertheless, the nonreciprocal quantum battery continues to exhibit a higher average charging power than its reciprocal counterpart. These results demonstrate that, although the DOPA parameters can significantly modulate the charging dynamics, the nonreciprocal energy-transfer mechanism maintains its advantage in charging power over the reciprocal configuration within the parameter regime considered.

\subsection{Optimization of the dissipation-channel asymmetry parameters}

\begin{figure}[ht]
\centerline{
\includegraphics[width=0.43\textwidth]{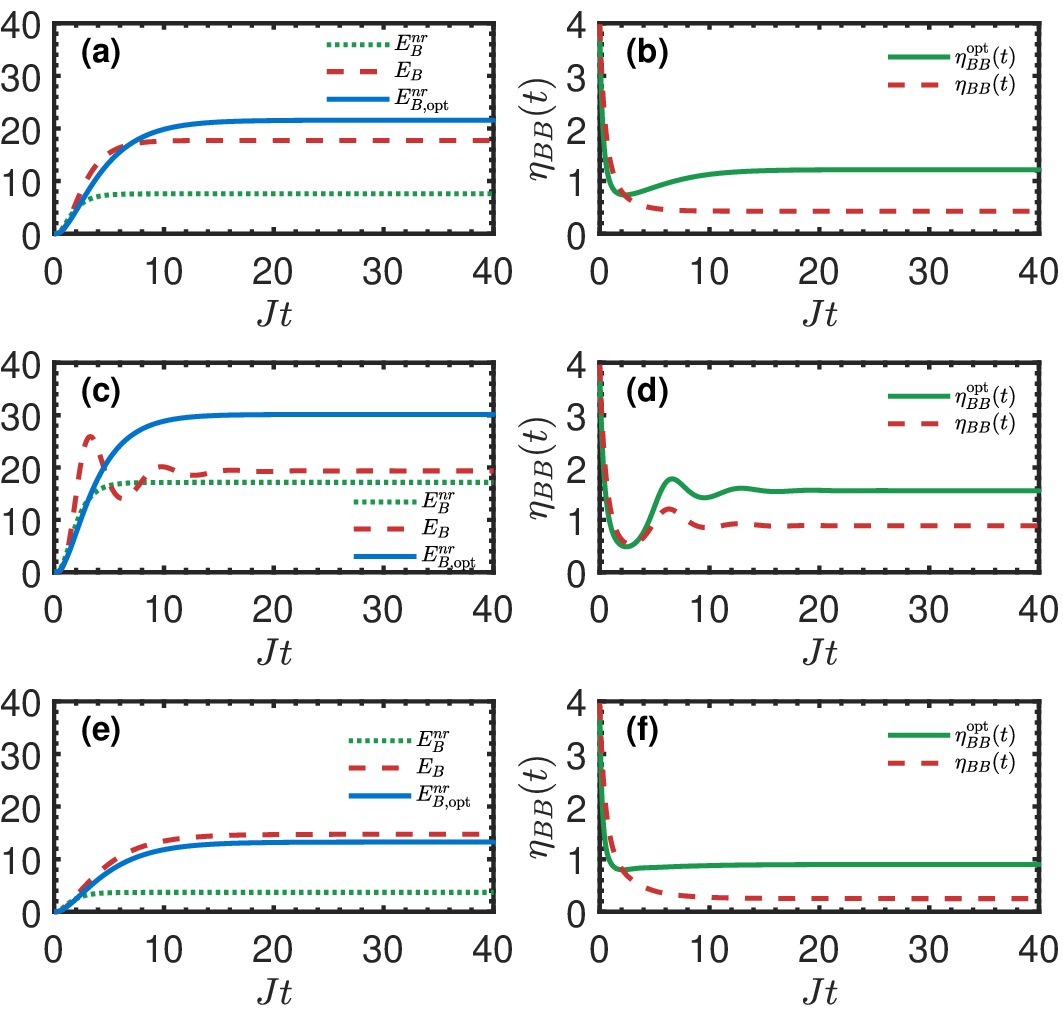}}
\caption{Comparison of the battery energies in the reciprocal 
    $E_B(t)$, nonreciprocal $E_B^{\text{nr}}(t)$, and optimized 
    nonreciprocal $E_{B,\text{opt}}^{\text{nr}}(t)$ regimes, shown 
    in (a)(c)(e), and the ratio $\eta_{BB}(t)$ versus its optimized 
    counterpart $\eta_{BB}^{\text{opt}}(t)$, shown in (b)(d)(f), 
    plotted against the scaled time $Jt$. 
    The DOPA nonlinear gain $g$, pump-field phase $\theta$, and 
    asymmetry parameter $\xi$ are set to: 
    (a) and (b) $g = 0$, $\theta = 0$, $\xi = 0.176$; 
    (c) and (d) $g = 0.02\omega$, $\theta = 0$, $\xi = 0.313$; 
    (e) and (f) $g = 0.02\omega$, $\theta = 3\pi/2$, $\xi = 0.221$. 
    The other parameters are ${\kappa_a}= 0.1{\omega}$, ${\kappa_b} = 0.003{\omega}$, ${\Gamma} =0.03{\omega}$, ${\Gamma _1}={\Gamma}/{\xi}$, ${\Gamma _2}={\xi}{\Gamma}$, ${\Gamma_3} =0.03{\omega}$, $f = 0.01\omega$, $\phi=0$,  and $|J| = 0.02\omega$.} \label{Two_energy_b_b_optb.eps}
\end{figure}

Analogous to the single-common-reservoir case, we optimize the parameters of the two-common-reservoir system by introducing an asymmetry parameter $\xi$, such that $\Gamma_1 = \Gamma/\xi$ and $\Gamma_2 = \Gamma\xi$. This parameterization scheme maintains a constant product of the system dissipation strengths ($\Gamma_1\Gamma_2 = \Gamma^2$), while allowing the asymmetry between the two dissipative channels to be continuously tuned via $\xi$. Substituting the above parameters into Eq.~\eqref{eq:EB_two_res}, taking the partial derivative with respect to $\xi$ and setting it to zero yields
\begin{equation}
    \frac{\partial E_B^{\text{nr}}(\infty)}{\partial \xi}\bigg|_{\xi = \xi_{\text{opt}}} = 0.
    \label{eq:opt_condition}
\end{equation}
Since the analytical solution of Eq.~\eqref{eq:opt_condition} is highly complex, we numerically determine the optimal parameter $\xi_{\text{opt}}$ and thereby establish the optimal dissipation configuration $(\Gamma_1^{\text{opt}},\, \Gamma_2^{\text{opt}})$. Substituting the obtained $\xi_{\text{opt}}$ into Eqs.~\eqref{Ei} and~\eqref{second12} yields the optimized nonreciprocal battery energy $E_{B,\text{opt}}^{\text{nr}}(t)$. Figures~\ref{Two_energy_b_b_optb.eps}(a)(c)(e) compare the battery energies across the three configurations. The results demonstrate that the optimized nonreciprocal battery $E_{B,\text{opt}}^{\text{nr}}(t)$ consistently exhibits the highest energy storage capacity. Specifically, Fig.~\ref{Two_energy_b_b_optb.eps}(a) illustrates that even in the absence of the DOPA, parameter optimization alone enables the nonreciprocal battery to achieve the maximum steady-state energy storage, whereas the unoptimized nonreciprocal battery stores less energy than its reciprocal counterpart. As depicted in Fig.~\ref{Two_energy_b_b_optb.eps}(c), upon introducing the DOPA ($g = 0.02\omega$), the steady-state battery energies of all systems are enhanced, with the optimized nonreciprocal battery maintaining its performance advantage. Furthermore,  Fig.~\ref{Two_energy_b_b_optb.eps}(e) reveals that shifting the DOPA phase $\theta$ from $0$ to $3\pi/2$ suppresses the battery energies in all systems. 
As shown in Figs.~\ref{Two_energy_b_b_optb.eps}(b)(d)(f), both $\eta_{BB}^{opt}$ and $\eta_{BB}$ grow with $g$ but are reduced when $\theta$ is tuned from $0$ to $3\pi/2$. Collectively, these findings indicate that the nonlinear gain $g$ and the pump-field phase $\theta$ of the DOPA offer robust tuning capabilities for the two-common-reservoir coupled quantum battery. This corroborates the conclusions drawn from the single-common-reservoir case and further demonstrates the robustness of the DOPA-enhanced nonreciprocal charging mechanism.
 
\section{Conclusions and Discussions}

In summary, we have investigated a reservoir-engineered quantum battery in which a  charger and battery are coupled through coherent and dissipative interactions, while the charger is driven simultaneously by a coherent field and a DOPA. By appropriately balancing the coherent and reservoir-induced dissipative couplings, the backaction of the battery on the charger can be suppressed, producing a nonreciprocal channel for directional energy transfer. We considered both single- and two-common-reservoir configurations and derived the corresponding dynamical equations and steady-state battery energies.
For the single-common-reservoir configuration, the DOPA provides an additional two-photon parametric-amplification channel that can markedly enhance both the charger and battery energies in the stable below-threshold regime. The nonlinear gain controls the strength of this enhancement, whereas the pump phase determines the interference between the coherent drive and the parametric process. Consequently, the stored energy and average charging power can be increased or suppressed by tuning the DOPA phase. Comparison with the reciprocal configuration shows that reservoir-induced nonreciprocity provides a clear advantage in the charging power over the parameter regimes studied, while the relative energy enhancement depends on the DOPA and dissipation parameters. Moreover, introducing an asymmetry parameter for the dissipative couplings and optimizing it numerically yields a further increase in the steady-state battery energy beyond the unoptimized configuration.
 We have further shown that these main features persist when the charger and battery are coupled to two common reservoirs. The DOPA again enhances the absolute stored energy and charging power for suitable gain and phase, and the pump phase provides an effective means of controlling the charging dynamics. Optimization of the dissipative-channel asymmetry is particularly important in this configuration: the optimized nonreciprocal battery achieves a larger steady-state energy than both the reciprocal and unoptimized nonreciprocal configurations for the cases examined. These results demonstrate that parametric amplification and reservoir-engineered nonreciprocity provide complementary control mechanisms for quantum-battery charging. Their combination offers a tunable route for enhancing energy storage and charging power in open bosonic quantum systems.  Several extensions of the present framework are worth exploring. First, the common reservoirs considered here can be generalized beyond the Markovian approximation to structured non-Markovian environments with finite correlation times. In such a regime, reservoir memory and the associated energy backflow may substantially modify the reservoir-mediated dissipative interaction and, consequently, the nonreciprocal charging dynamics. It would therefore be interesting to investigate how non-Markovian memory effects interplay with DOPA-induced parametric amplification and whether reservoir engineering can exploit such memory effects to further improve or dynamically control the stored energy and charging power. Second, the present two-mode charger–battery architecture can be extended to multi-cell quantum-battery systems, in which a charger is coupled to multiple battery units through common reservoirs. Such a generalization would enable the study of collective charging and reservoir-mediated correlations among different battery cells, as well as the scaling of the stored energy and charging power with the number of battery units. These extensions would broaden the present framework from Markovian few-mode settings toward non-Markovian and scalable quantum-battery architectures.
\section*{ACKNOWLEDGMENTS}
This work was supported by Science and Technology Development Plan Project of Jilin 
Province (Grant No. 20250102007JC), National Natural Science Foundation of China under Grant No. 12274064, and the Research Start-up Funds of Shenyang Normal University (Grants No. BS202516).

\section*{DATA AVAILABILITY}
The data that support the findings of this article are not publicly available. The data are available from the authors upon reasonable request.


\end{document}